\documentclass[pdflatex,sn-mathphys-num]{sn-jnl}

\usepackage{graphicx}%
\usepackage{multirow}%
\usepackage{amsmath,amssymb,amsfonts}%
\usepackage{amsthm}%
\usepackage{mathrsfs}%
\usepackage[title]{appendix}%
\usepackage{textcomp}%
\usepackage{manyfoot}%
\usepackage{booktabs}%
\usepackage{algorithm}%
\usepackage{algorithmicx}%
\usepackage{algpseudocode}%
\usepackage{listings}%
\usepackage{float}
\usepackage{placeins}
\usepackage{fancyhdr}  
\usepackage[utf8]{inputenc}
\usepackage[x11names]{xcolor}
\usepackage{pifont}             
\usepackage{makecell}
\usepackage{etoolbox}
\usepackage{titlesec}
\usepackage{enumitem}
\usepackage{hyperref}

\definecolor{mygreen}{rgb}{0.1,0.6,0.1}
\newcommand{\cmark}{\textcolor{mygreen}{\ding{51}}}  
\newcommand{\xmark}{\textcolor{red}{\ding{55}}}  

\theoremstyle{thmstyleone}%
\theoremstyle{thmstyletwo}%

\theoremstyle{thmstylethree}%

\begin{document}

\title[Article Title]{PocketVina Enables Scalable and Highly Accurate Physically Valid Docking through Multi-Pocket Conditioning}


\author*[1]{\fnm{Ahmet} \sur{Sarigun}}\email{Ahmet.Sariguen@mdc-berlin.de}

\author[1]{\fnm{Bora} \sur{Uyar}}

\author[1]{\fnm{Vedran} \sur{Franke}}

\author*[1]{\fnm{Altuna} \sur{Akalin}}\email{Altuna.Akalin@mdc-berlin.de}

\affil[1]{
\fontsize{9}{11}\selectfont
\orgdiv{Bioinformatics and Omics Data Science Platform}, \orgname{Max Delbruck Center for Molecular Medicine - Berlin Institute for Molecular Systems Biology},
\orgaddress{\street{Hannoversche Str. 28}, \city{Berlin}, \postcode{10115}, \state{Berlin}, \country{Germany}}}


\abstract{Sampling physically valid ligand-binding poses remains a major challenge in molecular docking, particularly for unseen or structurally diverse targets. We introduce PocketVina, a fast and memory-efficient, search-based docking framework that combines pocket prediction with systematic multi-pocket exploration. We evaluate PocketVina across four established benchmarks—PDBbind2020 (timesplit and unseen), DockGen, Astex, and PoseBusters—and observe consistently strong performance in sampling physically valid docking poses. PocketVina achieves state-of-the-art performance when jointly considering ligand r.m.s.d. and physical validity (PB-valid), while remaining competitive with deep learning–based approaches in terms of r.m.s.d. alone, particularly on structurally diverse and previously unseen targets. PocketVina also maintains state-of-the-art physically valid docking accuracy across ligands with varying degrees of flexibility. We further introduce TargetDock-AI, a benchmarking dataset we curated, consisting of over 500,000 protein–ligand pairs, and a partition of the dataset labeled with PubChem activity annotations. On this large-scale dataset, PocketVina successfully discriminates active from inactive targets, outperforming a deep learning baseline while requiring significantly less GPU memory and runtime. PocketVina offers a robust and scalable docking strategy that requires no task-specific training and runs efficiently on standard GPUs, making it well-suited for high-throughput virtual screening and structure-based drug discovery.}


%
%
%

\keywords{Molecular Docking, Search-based algorithm, Multi-Pocket Conditioned Docking}



\maketitle

\section*{Introduction}\label{sec1}

\vspace{1em}

\hspace*{1.0em} One of the first steps in drug discovery is the in silico screening of candidate molecules against target proteins. Molecular docking enables this process at scale, helping identify bioactive compounds from vast chemical libraries and thereby accelerating early-stage discovery pipelines. Moreover, recent annotations from the Drug Repurposing Hub, nearly 200 of 2,427 FDA-approved drugs lack a known mechanism of action \cite{corsello2017drug}, underscoring the need for efficient docking tools capable of screening against large target spaces. The efficacy of a drug is closely tied to how strongly and specifically it binds to its target protein, as well as its off-target effects. Rapid and accurate prediction of protein-ligand complexes is therefore essential for identifying promising candidates in virtual screening campaigns.

\vspace{1em}

The process of predicting how small molecules bind to a protein - often referred to as protein-ligand molecular docking - is highly complex due to the dynamic and multidimensional nature of protein-ligand interactions. This process generally consists of two steps: sampling the poses and scoring these poses \cite{tran2023practical}. Traditional methods such as AutoDock Vina \cite{trott2010autodock, eberhardt2021autodock}, Glide \cite{friesner2004glide}, and Gold \cite{verdonk2003improved} use heuristic or exhaustive search algorithms to discover potential ligand conformations. Simple scoring functions in these methods estimate the binding affinity of docked poses in energy terms. Several research groups developed SMINA \cite{koes2013lessons}, GNINA \cite{ragoza2017protein}, DeepDock \cite{mendez2021geometric}, and other machine learning-based scoring approaches \cite{cao2024generic, li2021machine} to improve the performance of these search techniques.

\vspace{1em}

Deep learning (DL) models have made attempts to unravel the complex nature of protein-ligand interactions \cite{xiong2022featurization, jumper2021highly, sadybekov2023computational, lin2023evolutionary, watson2023novo}. The first generation of DL–based docking methods, including EquiBind \cite{stark2202equibind}, TANKBind \cite{lu2022tankbind}, E3Bind \cite{zhang2022e3bind}, Uni-Mol \cite{zhou2023uni}, and KarmaDock \cite{zhang2023efficient}, formulated the problem as a regression problem. Although these approaches are relatively fast, some of them achieved only modest improvements in geometric accuracy—as measured by root-mean-square deviation (r.m.s.d.)—over traditional methods. On the other hand, Corso \emph{et al.} have considered molecular docking within the framework of deep generative modeling and aimed to learn the distribution of possible ligand poses with DiffDock \cite{corso2022diffdock}. They then introduced DiffDock-L \cite{corso2024deep}, which scaled the model size and improved r.m.s.d. within threshold accuracy (predicted ligand pose is considered geometrically accurate when its r.m.s.d. relative to the crystal ligand structure is below 2 Å \cite{alhossary2015fast}) in blind docking.

\vspace{1em}

The field continues to face unresolved challenges. Many deep learning models treat the entire protein as a potential binding surface, even though practical applications often identify the ligand binding site through prior knowledge or experimental data \cite{lyu2019ultra, bender2021practical, ackloo2022cache, kaplan2022bespoke, zhang2024advancing}. According to Bender \emph{et al.}, a suitable target site is the starting point for every structure-based campaign \cite{bender2021practical}. Although blind docking aims to solve pocket detection and pocket-conditional docking in a single step, the recent study by Yu \emph{et al.} showed that existing DL models, such as DiffDock, cannot overcome these two-step traditional approaches \cite{yu2023deep}. SurfDock \cite{cao2025surfdock} represents the latest attempt, using a diffusion process that conditions on the protein pocket and starts from a randomly chosen initial ligand conformation. SurfDock uses protein surface features, residue-level structural details, and pre-trained sequence information to build its surface node representation and achieves high r.m.s.d. within threshold success rates across several benchmarks.

\vspace{1em}

While many poses predicted by DL models were technically successful in terms of the r.m.s.d. within threshold, they are subject to physical inconsistencies. These inconsistencies can manifest as intermolecular steric clashes or unrealistic bond lengths and angles \cite{buttenschoen2024posebusters}. Considering the existing limitations, studies have focused on developing more robust metrics to assess the rationality of the generated poses \cite{buttenschoen2024posebusters, harris2023benchmarking}. Deane \emph{et al.} introduced the PoseBusters tool that analyzes poses based on physical and chemical consistency \cite{buttenschoen2024posebusters}. Their results show that DL methods are not superior to traditional methods in producing plausible poses and that the performance of all DL models degrades significantly for proteins with less than 30\% sequence similarity to the training set. This suggests that current DL algorithms have difficulty not only in generating biophysically consistent complex ligand structures, but also in generalizing to novel proteins. The main reason is the quality of the data used for training DL models. Analysis of the structures on the PDBBind dataset \cite{wang2004pdbbind} revealed that they contain a highly fraction of implausible structures \cite{sarigun2024compassdock, durairaj2024plinder}. In conclusion, it appears that over-reliance on r.m.s.d. in pose assessment is increasingly inadequate; it fails to capture the subtleties of molecular interactions and the physical limitations of binding poses.

\vspace{1em}

Protein pocket detection efforts have played a critical role for the success of pocket conditional protein-ligand docking, and many methods have been developed specifically for the identification of ligand binding sites \cite{ngan2012ftsite, tan2013depth}. Among geometric-based tools, Fpocket \cite{le2009fpocket} offers fast and independent pocket prediction by filtering and clustering alpha spheres detected through Voronoi tessellation; although it is widely used in large-scale applications, it may not always find the known binding site at the top of the predicted pockets. SiteHound \cite{ghersi2009easymifs}, an energy-based approach, identifies pockets by calculating interaction energies through a probe applied to grid points positioned around the protein surface. Consensus-based MetaPocket 2.0 \cite{zhang2011identification} combines the top three predictions from each of eight different algorithms to produce results that outperform individual methods. DeepSite \cite{jimenez2017deepsite}, a DL-based method, performs highly accurate pocket detection by processing three-dimensional space into voxelized representations for different atom types and analyzing them with deep convolutional neural networks. P2Rank \cite{krivak2018p2rank, polak2025prankweb}, which uses a random forest classifier, evaluates statistical properties of surface nodes such as atom type, neighborhood and curvature as attributes and has become a preferred method in both open source and commercial environments due to its fast execution.

\vspace{1em}

Blind docking (docking without pocket detection) lacks both time and resource efficiency by having to find the ligand binding site and optimize conformations over the entire protein surface, whereas single-pocket conditional docking is fast and accurate in binding to the defined site but has limited generalizability. An early work \cite{chen2014docking} demonstrated a limited increase in success by docking multiple protein pocket regions simultaneously; however, since neither reliable pocket detection methods nor GPU-accelerated docking were widespread at the time, this idea was not sufficiently adopted.


\vspace{1em}

To overcome this dichotomy, we reinterpreted classical docking and developed PocketVina, a hybrid multi-pocket-conditioned docking approach that provides both fast pocket classification with P2Rank followed by physically valid accurate pose sampling with QuickVina 2-GPU 2.1. QuickVina 2-GPU 2.1 \cite{tang2024vina} is an optimized version of the AutoDock Vina family \cite{trott2010autodock, eberhardt2021autodock, tang2022accelerating, ding2023vina} for high-throughput virtual screening (HTVS), further improving the speed of Vina-GPU 2.0 \cite{ding2023vina} with the RILC-BFGS (Reduced Iteration and Low Complexity BFGS) algorithm to a few tens of milliseconds per dock. PocketVina achieves large-scale docking throughput within just a few days, even when executed on standard GPU hardware (e.g., NVIDIA Tesla P40 with ~6 GB VRAM). This demonstrates both the scalability and accessibility of the approach, making it particularly well suited for high-throughput virtual screening applications. PocketVina performs exhaustive sampling across multiple detected pocket centers using Vina's GPU acceleration and repeats the optimization for multiple pockets, resulting in more rational pose estimates. In several benchmarks - PDBbind2020 \cite{wang2004pdbbind, stark2202equibind}, Astex Diverse Set \cite{hartshorn2007diverse}, PoseBusters benchmark set \cite{buttenschoen2024posebusters}, and DockGen \cite{corso2024deep} - PocketVina showed the highest physically-valid success rates for rigid ligand docking compared to both traditional and DL-based models. We curated the TargetDock-AI benchmarking dataset to evaluate how well docking programs’ own scoring functions can distinguish active from inactive drug–target pairs. In this benchmark, PocketVina clearly outperformed the DL-based method by achieving accurate discrimination with minimal false positives. On top of this, PocketVina's runtime efficiency provides a valuable contribution to the structure-based drug design (SBDD) community in real-world high-throughput virtual screening scenarios.

\section*{Results}\label{sec2}

\subsection*{Method Overview}\label{sec2.1}

\begin{figure}[h]
\centering
\includegraphics[width=0.99\textwidth]{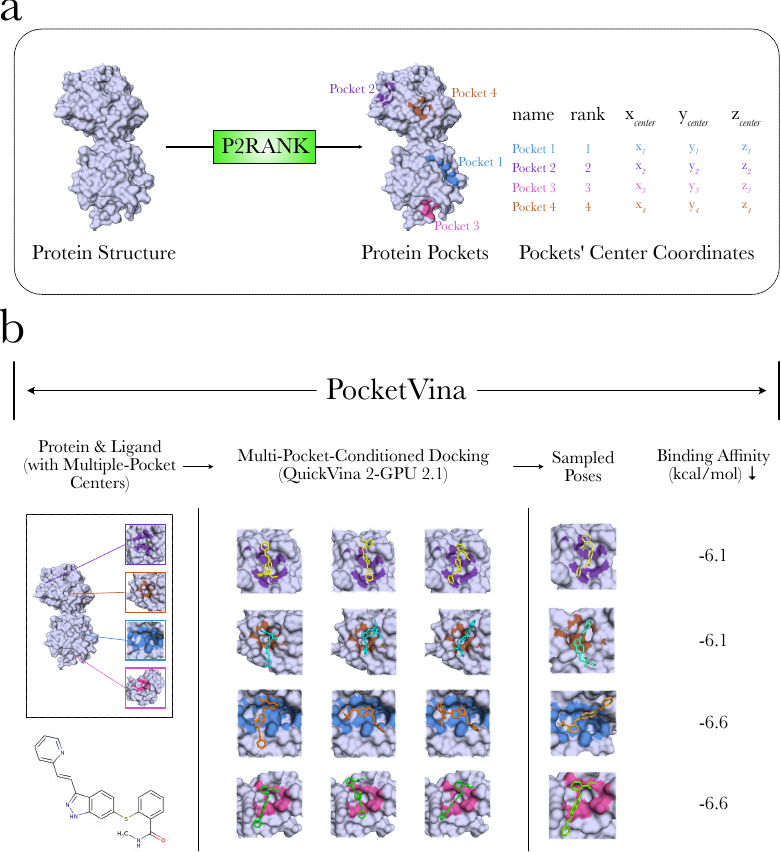}
\caption{\textbar{} \textbf{The Overview of PocketVina.} 
\newline
\textbf{(a)} Classification of protein pocket centers with P2Rank. P2Rank calculates the ligandability scores of proteins with SAS points and Random Forest to find pocket scores and rank the pockets. After classifying the pockets, it calculates pocket centers. \textbf{(b)} Illustration of PocketVina. QuickVina 2-GPU 2.1 performs molecular docking to multiple pocket centers. Scoring function of QuickVina 2-GPU 2.1 calculates the binding affinity energy of the docked molecules.}
\label{figure_1}
\end{figure}

\vspace{1em}

\hspace*{1.0em} Our protein-ligand docking workflow, PocketVina, operates in two stages: 1) prediction of potential protein pocket centers using P2Rank, a fast and accurate machine learning algorithm that identifies ligandable regions on the protein's solvent-accessible surface (SAS); and 2) molecular docking at each predicted pocket center using QuickVina 2-GPU 2.1, an accelerated version of AutoDock Vina optimized for GPU-based virtual screening.

\vspace{1em}

P2Rank rapidly analyzes protein structure to identify and rank putative ligand-binding pockets. It classifies points on the SAS of the protein and clusters high-scoring regions to produce discrete pocket center predictions, which PocketVina uses as independent docking targets (Figure \ref{figure_1}a).

\vspace{1em}

QuickVina 2-GPU 2.1 then independently processes each pocket center and performs parallelized docking using GPU acceleration. It applies an efficient optimization algorithm to generate and refine ligand conformations, calculating each pose’s binding affinity (Figure \ref{figure_1}b).

\vspace{1em}

The workflow begins with the identification of binding pockets and conversion of the input protein into the Vina-compatible PDBQT format. We use RDKit \cite{landrum2006rdkit} and Open Babel \cite{o2011open} to process ligand input—provided as a simplified molecular-input line-entry system (SMILES) \cite{weininger1988smiles} string or 2D representation—and generate a randomized 3D conformation, which we then convert to PDBQT format. Finally, QuickVina 2-GPU 2.1 then performs a search-based optimization across the identified pocket centers to refine the initial pose and generate a user-defined number of docked conformations.

\vspace{1em}

We conducted extensive benchmarking to evaluate PocketVina’s performance. Across diverse datasets, the method consistently demonstrated strong sampling capability, with results largely independent of the scoring function.

\subsection*{PocketVina achieves physically valid docking accuracy beyond  state-of-the-art methods}\label{sec2.2}

\vspace{1em}

\hspace*{1.0em} Table \ref{table_1} shows that, PocketVina achieves a significantly higher success rate on the PDBbind2020 time-split test set (363 complexes), when evaluated using the PoseBusters (PB) tool, outperforming both deep learning (DL) and traditional docking models. PocketVina’s results reflect the most accurate pose among the sampled conformations, whereas literature-reported results follow the respective scoring protocols of each method. Compared to traditional methods such as GNINA and GLIDE-SP, as well as DL-based approaches such as DiffDock and SurfDock, PocketVina achieves a significantly higher success rate in sampling physically valid docking poses.  Notably, 50.96\% of the poses by PocketVina have an $\text{r.m.s.d.} < 2\,\text{\AA}$ and PB-valid criteria (Supplementary \ref{supp:S3}), while the closest competitor remains at 41.41\% (Table \ref{table_1}). This finding supports the view proposed by Buttenschoen \emph{et al.} \cite{buttenschoen2024posebusters}, that while DL models can learn the geometry of the rigid ligand structure, they are unable to provide physically valid docking poses, whereas existing traditional methods provide more accurate predictions. Although the r.m.s.d. metric does not provide direct information about physical validity, PocketVina achieves competitive results at both $\text{r.m.s.d.} < 1\,\text{\AA}$ and $\text{r.m.s.d.} < 2\,\text{\AA}$ thresholds, with median (Å) values also showing the best performance. Supplementary Figure \ref{Supp_figure_1}a-b, shows that PocketVina's r.m.s.d. distributions are very competitive compared to other methods (Supplementary \ref{supp:S4}). In the PDBbind2020 test set, PocketVina outperforms all methods when compared on new proteins (144 complexes) that are not included in the training data of DL models, not only in the $\text{r.m.s.d.} < 2\,\text{\AA}$ \& PB-valid threshold metrics, but also the poses that meet the $\text{r.m.s.d.} < 1\,\text{\AA}$ and $\text{r.m.s.d.} < 2\,\text{\AA}$ thresholds metrics. Since this test set does not have “rigid overlap” \cite{su2020tapping} with the proteins used in the training of the DL models - i.e., it does not contain the same structures - these results point to PocketVina's superior generalization ability against novel proteins. Furthermore, we restricted the benchmark to rigid docking methods to ensure fast post-processing in high-throughput virtual screening scenarios, while also providing a fair assessment of the accuracy achievable by rigid models alone. 


\vspace{1em}

\begin{table}[h]
\definecolor{mygreen}{rgb}{0.1, 0.6, 0.1}
\caption{\textbar{} Comparative Analysis of Docking Performances on PDBbind2020 Dataset}\label{table_1}
\begin{tabular*}{\textwidth}{@{\extracolsep\fill}lccccc}
\toprule
\textbf{Model Type} & \textbf{Method} & \textbf{\% $<$1\AA\ } & \textbf{\% $<$2\AA\ } & \textbf{Med($\text{\AA}$)} & \textbf{\% $<$2\AA\ \& PB} \\
\midrule
\textbf{DL}  & EquiBind & - & 5.5$\pm$1.2 & 6.2$\pm$0.3 & - \\
\textbf{DL}  & TANKBind & 2.66$\pm$0.26 & 18.18$\pm$0.6 & 4.2$\pm$0.05 & - \\
\textbf{DL}  & E3Bind & - & 25.6 & 7.2 & - \\
\textbf{DL}  & KarmaDock & - & \textcolor{mygreen}{56.2} & - & - \\
\textbf{DL}  & DiffDock(Pocket) & - & 51.8 & 2.0 & - \\
\textbf{DL}  & DiffDock & 15.15 & 36.09 & 3.35 & 15.43 \\
\textbf{DL}  & DiffDock-L & 19.07$\pm$0.57 & 40.74$\pm$1.25 & 2.88$\pm$0.18 & 21.85$\pm$0.52 \\
\textbf{DL}  & SurfDock(unmin) & \textcolor{red}{40.96$\pm$0.34} & \textcolor{red}{68.41$\pm$0.26} & \textcolor{blue}{1.18$\pm$0.00} & 36.46$\pm$0.26 \\
\textbf{Classical}  & Uni-Dock & \textcolor{mygreen}{32.51$\pm$0.39} & 50.69$\pm$0.59 & \textcolor{mygreen}{1.89$\pm$0.04} & 20.29$\pm$0.94 \\
\textbf{Classical}  & Glide-SP & 17.36$\pm$0.00 & 44.63$\pm$0.00 & 2.27$\pm$0.00 & \textcolor{mygreen}{38.57$\pm$0.00} \\
\textbf{Classical}  & GNINA & 21.12$\pm$0.26 & 43.62$\pm$1.06 & 2.45$\pm$0.07 & \textcolor{blue}{41.41$\pm$1.13} \\
\textbf{Classical}  & SMINA & 18.73$\pm$0.00 & 31.68$\pm$0.00 & 3.99$\pm$0.00 & 28.37$\pm$0.00 \\
\textbf{Classical}  & Vina & 18.32$\pm$0.02 & 36.64$\pm$0.05 & 3.42$\pm$0.01 & 32.87$\pm$0.91 \\
\textbf{Hybrid}  & PocketVina & \textcolor{blue}{39.39} & \textcolor{blue}{59.22} & \textcolor{red}{1.15} & \textcolor{red}{50.96} \\
\midrule
\multicolumn{6}{l}{Performance on unseen proteins in PDBbind2020 time-split test set (144 complexes)} \\
\midrule
\textbf{DL}  & DiffDock & 4.86 & 18.75 & 5.81 & 5.56 \\
\textbf{DL}  & DiffDock-L & 5.36$\pm$0.33 & 26.34$\pm$1.65 & 4.67$\pm$0.52 & 7.83$\pm$0.66 \\
\textbf{DL}  & SurfDock(unmin) & \textcolor{blue}{32.87$\pm$0.65} & \textcolor{blue}{60.88$\pm$0.33} & \textcolor{blue}{1.51$\pm$0.01} & 30.79$\pm$0.33 \\
\textbf{Classical}  & Uni-Dock & \textcolor{mygreen}{32.64$\pm$1.50} & 44.91$\pm$0.65 & 2.55$\pm$0.02 & 18.06$\pm$1.50 \\
\textbf{Classical}  & Glide-SP & 16.67$\pm$0.00 &  \textcolor{mygreen}{46.53$\pm$0.00} & \textcolor{mygreen}{2.13$\pm$0.00} & \textcolor{mygreen}{35.42$\pm$0.00} \\
\textbf{Classical}  & GNINA & 16.67$\pm$0.00 & 38.43$\pm$1.31 & 2.75$\pm$0.18 & \textcolor{blue}{36.57$\pm$0.87} \\
\textbf{Classical}  & SMINA & 11.81$\pm$0.00 & 27.08$\pm$0.00 & 4.32$\pm$0.00 & 24.31$\pm$0.00 \\
\textbf{Classical}  & Vina & 10.41$\pm$0.00 & 25.69$\pm$0.08 & 4.25$\pm$0.02 & 23.61$\pm$0.98 \\
\textbf{Hybrid}  & PocketVina & \textcolor{red}{40.97} & \textcolor{red}{61.81} & \textcolor{red}{1.01} & \textcolor{red}{52.08} \\
\bottomrule
\end{tabular*}
\footnotetext{Note: EquiBind, TANKBind, E3Bind and Uni-Dock results are taken from Ke \emph{et al.} \cite{yu2023deep}; KarmaDock from its original paper \cite{zhang2023efficient}. DiffDock’s blind-docking performance was obtained from the open-source repository maintained by Ke \emph{et al.} \cite{yu2023deep}(\url{https://github.com/pkuyyj/Blind_docking}), while its pocket-conditioned results follow Huang \emph{et al.} \cite{huang2024re}. Glide SP, GNINA, SMINA, Vina, DiffDock-L and SurfDock were obtained from SurfDock \cite{cao2025surfdock}. The top result is shown in \textcolor{red}{red}, second in \textcolor{blue}{blue}, third in \textcolor{mygreen}{green}. We compare all methods on the PDBbind2020 time-split test set and on novel protein targets using metrics for r.m.s.d. $<$ 1 Å/2 Å, median r.m.s.d., and a PB-valid score (poses passing all PoseBusters tests). For 363 complexes, PocketVina generated poses for 304 out of 363 protein–ligand pairs. To ensure a fair comparison, all success rates (\% $<$ 1 Å, \% $<$ 2 Å, and \% $<$ 2 Å \& PB-valid) were normalized against the full set of 363 pairs. The median r.m.s.d. (Med Å) is reported over the 304 pairs for which PocketVina produced results. For 144 complexes (unseen proteins in PDBbind2020 time-split test set), PocketVina generated poses for 120 out of 144 protein–ligand pairs. To ensure a fair comparison, all success rates (\% $<$ 1 Å, \% $<$ 2 Å, and \% $<$ 2 Å \& PB-valid) were normalized against the full set of 144 pairs. The median r.m.s.d. (Med Å) is reported over the 120 pairs for which PocketVina produced results. PocketVina results reflect the best r.m.s.d. values among all sampled poses, whereas literature-reported results correspond to poses selected by each method’s scoring function.}
\end{table}

\vspace{1em}

\begin{figure}[h]
\centering
\includegraphics[width=0.99\textwidth]{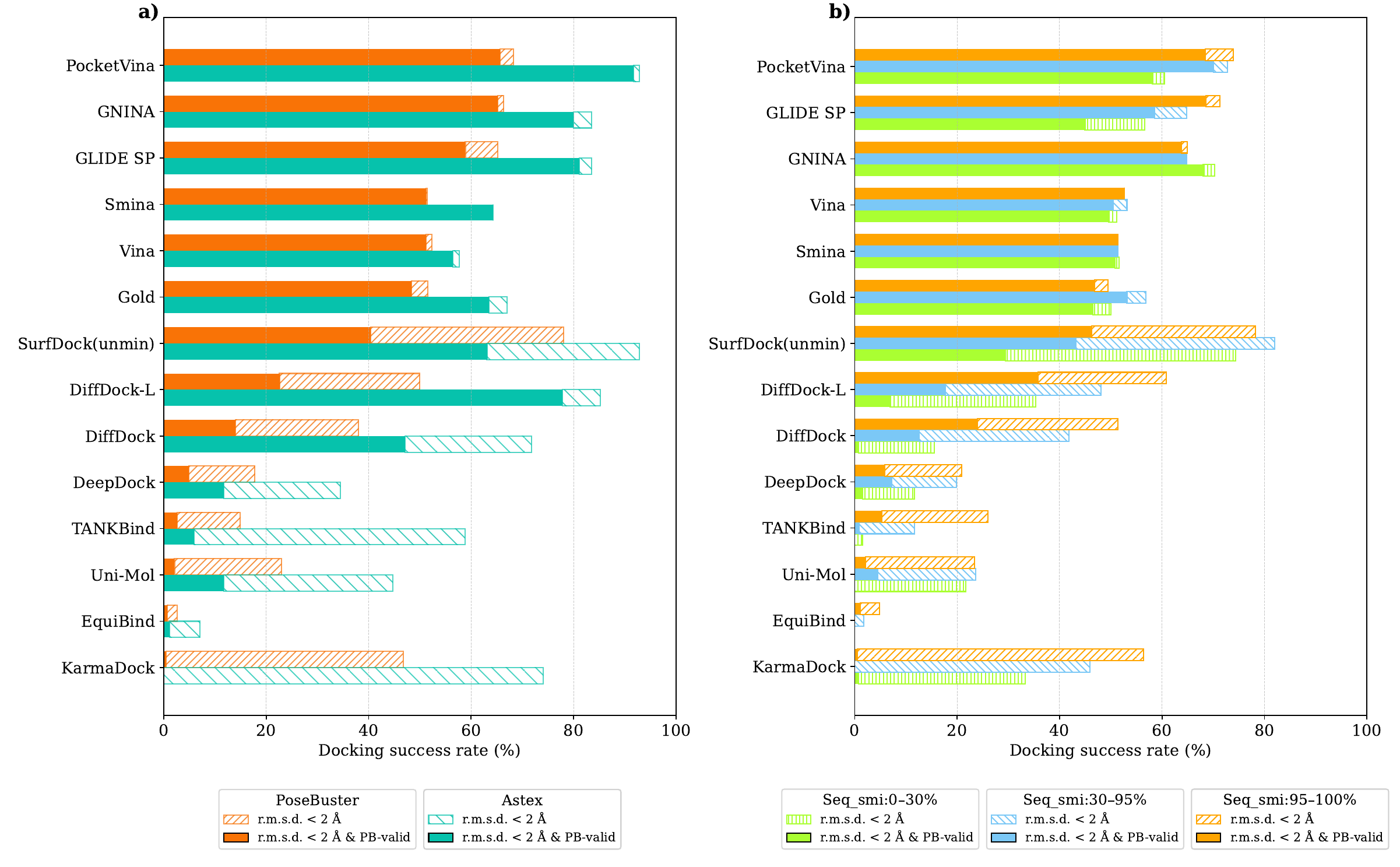}
\caption{\textbar{} \textbf{Comparative performance of docking methods across benchmarks.} 
\newline
\textbf{(a)} Comparison of docking-method performance on two benchmarks: the Astex Diverse set (85 cases) and the PoseBusters set (428 cases). Striped bars show the fraction of poses with r.m.s.d. $\leq$ 2 Å; solid bars show those that additionally pass the PoseBuster (PB-valid) criteria—i.e., satisfy key biophysical restraints. \textbf{(b)} Docking-method performance on the PoseBusters benchmark, stratified by sequence similarity to PDBbind2020. Striped bars indicate the fraction of poses with r.m.s.d. $\leq$ 2 Å, while solid bars represent those that also meet PB-valid criteria.}
\begin{minipage}{1.0\textwidth}
\footnotesize Note: For Astex and PoseBuster datasets, the reported results for KarmaDock, EquiBind, Uni-Mol, TANKBind, DeepDock, DiffDock, DiffDock-L, SurfDock(unmin), Gold, Vina, Smina, GLIDE SP and GNINA are taken from SurfDock \cite{cao2025surfdock} paper. Because our evaluation covers only rigid-docking protocols, we have excluded any post-processed docking results for these methods. PocketVina results reflect the best r.m.s.d. values among all sampled poses, whereas literature-reported results correspond to poses selected by each method’s scoring function.
\end{minipage}
\label{figure_2}
\end{figure}

We further used both the Astex Diverse set and the PoseBusters benchmark set (Figure \ref{figure_2}a) to investigate PocketVina’s performance on small drug-like molecules. In these benchmarks, we compared the reliability and generalizability of the poses obtained by different methods. PocketVina significantly outperformed the other methods on both datasets, achieving a 65.65\% success rate on the r.m.s.d. $<$ 2 Å \& PB-valid criteria on the PoseBusters set and 90.58\% on the Astex Diverse set (solid bar graphs in Figure \ref{figure_2}a). Compared to other DL and traditional methods, PocketVina achieved the highest docking performance in terms of ligand validity. Notably, PocketVina’s results reflect the most accurate pose per complex, whereas literature-reported results for other methods follow their respective scoring protocols. Furthermore, the r.m.s.d. cumulative distributions in Supplementary Figure \ref{Supp_figure_1}c-d clearly show that PocketVina offers competitive results with both DL and traditional approaches on r.m.s.d. $<$ 1 Å and r.m.s.d. $<$ 2 Å thresholds (Supplementary \ref{supp:S4}). While SurfDock offers the best performance when considering the r.m.s.d. as the exclusive criteria of success, its performance deteriorates when considering the physical validity of the docking results in combination with the r.m.s.d. 

\vspace{1em}

\vspace{1em}

We also evaluated PocketVina on a set of PoseBusters from PDBbind2020 categorized by protein sequence similarity, as shown in Figure \ref{figure_2}b. The low-similarity subset, which is not expected to have “soft overlap” \cite{su2020tapping} with proteins in the training data, showed remarkable results. While all other DL methods significantly lost physically valid docking success rates on proteins with less than 30\% sequence similarity, PocketVina showed only a slight decrease. In addition, PocketVina still showed competitive results when compared to traditional methods, notably producing more accurate poses than Vina, because Vina doesn’t take advantage of considering multiple pocket centers.

\vspace{1em}

By providing consistent physically-valid poses even for proteins with low sequence similarity, PocketVina demonstrates its ability to generalize to novel proteins. This is vital when we focus on novel targets with no known ligands \cite{ackloo2022cache}. The robustness and adaptability demonstrated by PocketVina both reinforces its reliability and positions it as a valuable tool for high throughput virtual screening, where it is critical to accurately identify ligands suitable for novel protein targets.

\vspace{1em}

\begin{table}[h]
\definecolor{mygreen}{rgb}{0.1, 0.6, 0.1}
\caption{\textbar{} Comparative Analysis of Docking Performances on DockGen-full Dataset}\label{table_2}
\begin{tabular*}{\textwidth}{@{\extracolsep\fill}lccccc}
\toprule
\textbf{Model Type} & \textbf{Method} & \textbf{\% $<$1\AA\ } & \textbf{\% $<$2\AA\ } & \textbf{Med($\text{\AA}$)} & \textbf{\% $<$2\AA\ \& PB} \\
\midrule
\textbf{Classical}  & GNINA & - & 17.5 & 8.1 & - \\
\textbf{DL}  & DiffDock (10) & - & 7.1 & 6.8 & - \\
\textbf{DL}  & DiffDock (40) & - & 6.0 & 7.3 & - \\
\textbf{DL}  & DiffDock-L (10) & \textcolor{mygreen}{4.78$\pm$1.75} & \textcolor{mygreen}{22.16$\pm$0.98} & \textcolor{mygreen}{4.80$\pm$0.36} & \textcolor{mygreen}{4.19$\pm$0.85} \\
\textbf{DL}  & SurfDock(unmin) (40) & \textcolor{red}{41.52$\pm$0.83} & \textcolor{red}{71.73$\pm$0.55} & \textcolor{red}{1.13$\pm$0.02} & \textcolor{blue}{32.74$\pm$0.56} \\
\textbf{Hybrid}  & PocketVina & \textcolor{blue}{23.80} & \textcolor{blue}{44.97} & \textcolor{blue}{2.02} & \textcolor{red}{39.68} \\
\bottomrule
\end{tabular*}
DockGen-full comprises 189 complexes which do not share similar binding pockets within the training datasets in PDBbind2020.
\footnotetext{Note: The reported results for GNINA, DiffDock, DiffDock-L and SurfDock are taken from SurfDock \cite{cao2025surfdock} paper. Because our evaluation covers only rigid-docking protocols, we have excluded any post-processed docking results for these methods. On the DockGen-full benchmark, PocketVina generated poses for 171 out of 189 protein–ligand pairs. To ensure a fair comparison, all success rates (\% $<$ 1 Å, \% $<$ 2 Å, and \% $<$ 2 Å \& PB-valid) were normalized against the full set of 189 pairs. The median r.m.s.d. (Med Å) is reported over the 171 pairs for which PocketVina produced results. PocketVina results reflect the best r.m.s.d. values among all sampled poses, whereas literature-reported results correspond to poses selected by each method’s scoring function.}
\end{table}

To further evaluate the generalizability of the results obtained using our workflow, we conducted experiments (Table \ref{table_2}) to evaluate PocketVina's performance on a test dataset called DockGen \cite{corso2024deep}, which contains new binding sites that were not included in the training sets of DL models. Surprisingly, PocketVina outperformed both traditional and DL-based methods on the DockGen dataset, especially in physically valid pose accuracy. This finding highlights that multi-pocket conditioned docking using classical docking algorithms offers the most generalizable results while also offering competitive results with both traditional and DL approaches in benchmarks based on r.m.s.d. alone.

\subsection*{PocketVina outperforms in physically valid docking across ligand flexibilities}\label{sec2.3}

\vspace{1em}

\begin{figure}[!htbp]
\centering
\includegraphics[width=0.96\textwidth]{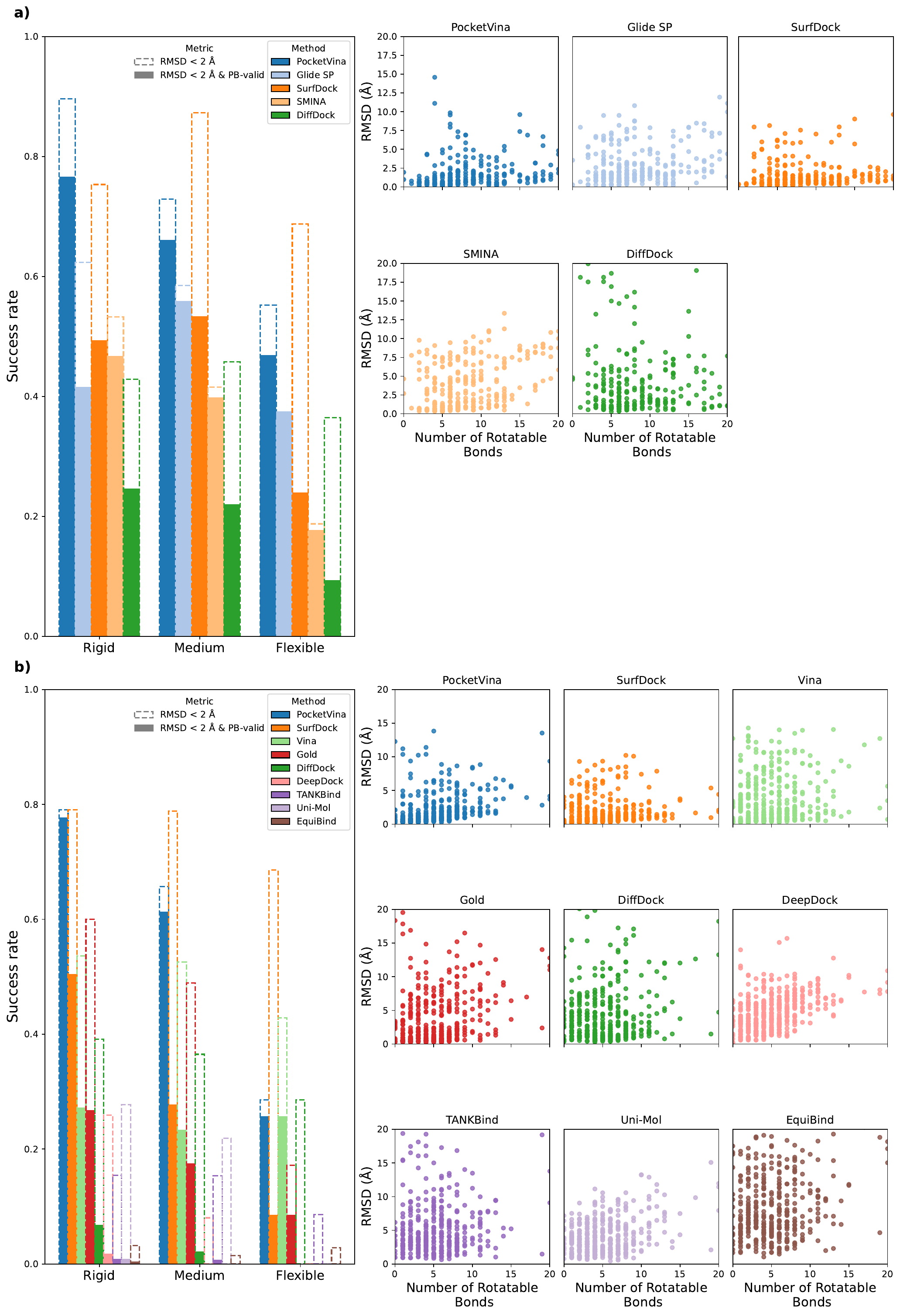}
\caption{\textbar{} \textbf{Physically-Valid performance of various docking methods across ligand flexibility categories} (rigid: $\leq$ 5 rotatable bonds; medium: 6–10; flexible: $>$ 10) 
\newline
\textbf{(a)} Effect of rotatable bonds on physically valid docking success rates and r.m.s.d. distribution in the PDBbind time-split set.\protect\footnotemark[1]
\textbf{(b)}Effect of rotatable bonds on physically valid docking success rates and r.m.s.d. distribution in the PoseBusters set.\protect\footnotemark[2]
}
\label{figure_3}
\end{figure}

\hspace*{1.0em} The flexibility of a small molecule, as measured by the number of rotatable bonds and the number of heavy atoms in a given ligand, presents a significant challenge, which prevents the  docking methods to achieve physically valid conformations \cite{cao2025surfdock}. To evaluate how different methods performed with such a challenge, we examined the effect of ligand flexibility on the physically valid docking success of different methods on the PDBbind2020 test set and the PoseBusters benchmark set. PocketVina offered the highest physically valid successful pose rates, consistently outperforming other methods across all flexibility and atom ranges (Figure \ref{figure_3}a-b and Extended Data Figure \ref{Ext_figure_1}a-b).
 
\footnotetext[1]{Fig. \ref{figure_3}-a Note: The reported results for Glide SP, SurfDock, SMINA and DiffDock on PDBBind time-split set are taken from SurfDock \cite{cao2025surfdock} paper. Because some PDB IDs lack PoseBusters evaluations, we included only those complexes with available PB analysis for each method—hence panel (a) comprises just n = 293 complexes. This also holds for Extended Data Fig. \ref{Ext_figure_1}a.}
\footnotetext[2]{Fig. \ref{figure_3}-b Note: Results for Vina, Gold, DiffDock, DeepDock, TANKBind, Uni-Mol and EquiBind on the PoseBuster benchmark are taken from the PoseBuster \cite{buttenschoen2024posebusters} paper, while SurfDock’s values are from \cite{cao2025surfdock}. Only complexes with available PoseBuster analysis for each method were included—hence panel (b) comprises n = 409 complexes. This also holds for Extended Data Fig. \ref{Ext_figure_1}b.}

\vspace{1em}

In both datasets, it surpassed both traditional and DL-based best methods for rigid and medium ligands. Especially for rigid ligands, it outperformed other methods by more than \%20 with a physically-valid successful pose rate of nearly 80\% in both datasets. For medium ligands, PocketVina achieved a success rate of more than 60\% in PDBbind2020, an increase of about 10\% compared to its competitors, while in the PoseBusters set the difference was over 30\%. With flexible ligands, PocketVina was still the top performer.

\vspace{1em}

Considering that most drugs and drug-like compounds have less than ten rotatable bonds, PocketVina is an ideal docking tool for such compounds \cite{wang2016comprehensive}. Furthermore, it continued to outperform the physically valid success rates of both open-source and commercial software in the “flexible” ligand group, as seen in Figure 3a-b. Although performance degraded with increasing flexibility in all models, PocketVina stood out as the most reliable method in both datasets.

\subsection*{Relationship between pocket ranks and binding affinity}\label{sec2.4}

\vspace{1em}

\hspace*{1.0em} As PocketVina depends on an initial pocket finding step, we investigated the relationship between pocket ranking and predicted binding affinity. We use all valid pockets predicted by P2Rank for docking; however, sampling more pockets does not necessarily lead to better binding affinity values. To better understand this relationship, we analyzed the distribution of Vina docking scores across different pocket rank bins.

\vspace{1em}

\begin{figure}[h]
\centering
\includegraphics[width=0.99\textwidth]{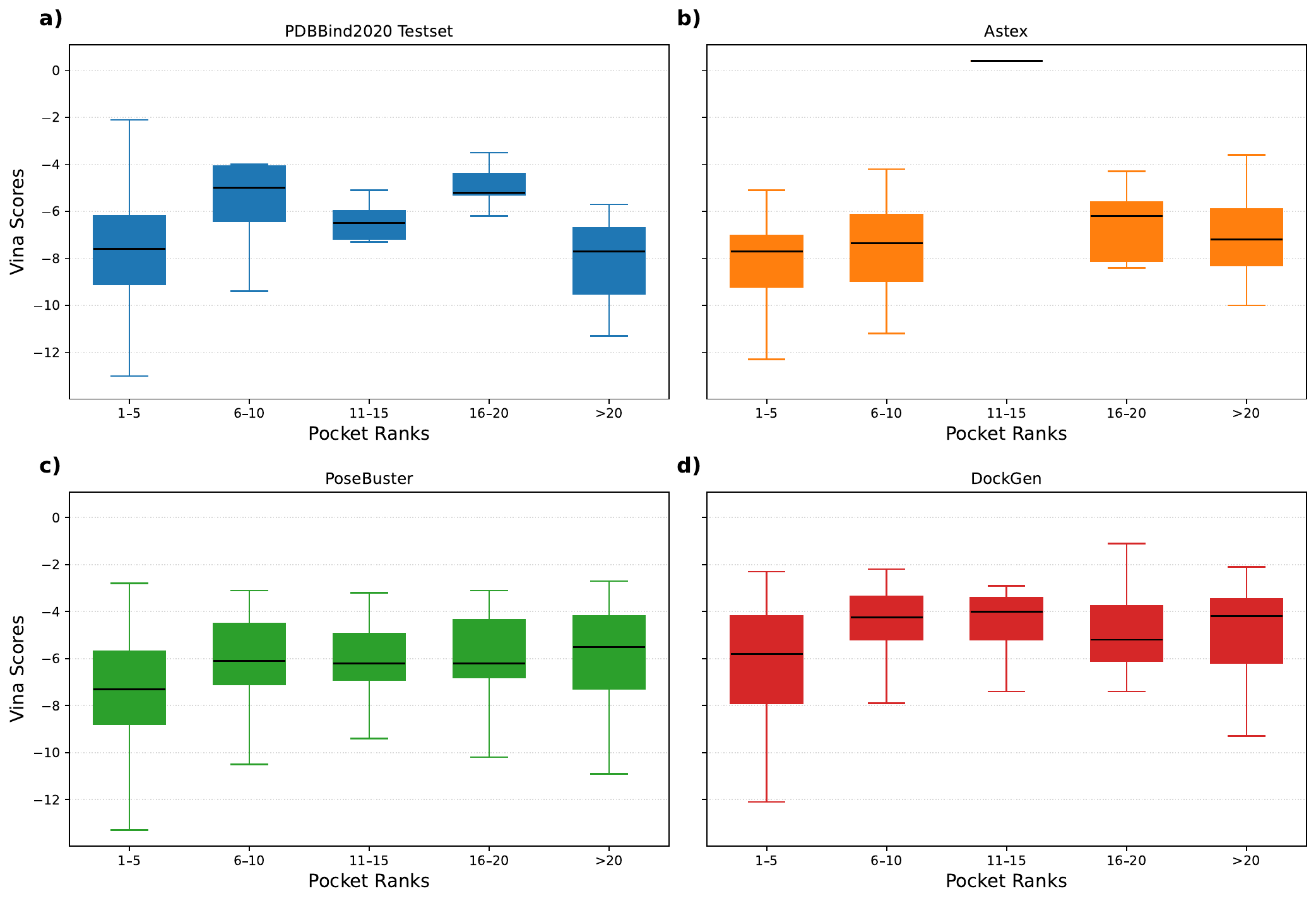}
\caption{\textbar{} \textbf{Relationship between pocket rank and predicted binding affinity across benchmark datasets.} 
\newline
Box plot distributions of Vina docking scores across grouped pocket ranks \textbf{(a)} PDBbind2020 time-split test set (363 complexes), \textbf{(b)} Astex Diverse set (85 complexes), \textbf{(c)} PoseBuster benchmark set (428 complexes), and \textbf{(d)} DockGen dataset (189 complexes). More negative Vina scores indicate stronger predicted binding affinity.}
\label{figure_4}
\end{figure}

For each protein–ligand complex, we grouped all pocket centers identified by P2Rank according to their rank (e.g., ranks 1–5, 6–10, etc.) and collected the Vina scores for each resulting pose. We then aggregated these scores across four benchmark datasets (PDBbind2020, Astex, PoseBuster, and DockGen) and plotted the distributions (Figure \ref{figure_4}). To ensure consistency with our evaluation pipeline, we extracted Vina scores only from the poses with the best r.m.s.d. per complex. As a result, the affinity values shown for each pocket rank reflect the best pose within that group, not necessarily the top-ranked one based on Vina scoring alone. While Vina scores provide a general indication of binding strength, they do not always align with the most biophysically accurate or correctly ranked pocket. 

\vspace{1em}

Across all datasets, we observed that higher-ranked pockets (particularly ranks 1–5) consistently yielded more favorable docking scores (i.e., lower Vina scores), indicating stronger predicted binding affinities. This trend was pronounced in all datasets (PDBbind2020, Astex, PoseBuster, and DockGen), where lower-ranked pockets showed more variable and generally weaker binding predictions. These results support the intuition that top-ranked pockets, as assigned by P2Rank, are more likely to correspond to biophysically relevant binding sites.

\vspace{1em}

Furthermore, since Vina scores remain the standard metric in high-throughput virtual screening pipelines, this analysis offers a practical insight: docking into the top five-ranked pockets is typically sufficient to identify strong candidates. This balance between computational efficiency and biophysical relevance makes PocketVina especially suitable for large-scale screening tasks \cite{ackloo2022cache}.

\subsection*{PocketVina rapidly distinguishes active from inactive drugs for given targets}\label{sec2.5}

\vspace{1em}

\hspace*{1.0em} To further validate PocketVina’s ability to identify biologically relevant interactions, we constructed a benchmark dataset, TargetDock-AI (AI = Active/Inactive), designed to evaluate whether docking-derived metrics can distinguish active from inactive drug–target complexes in the absence of task-specific training data.

\vspace{1em}

We curated this dataset for 237 proteins that are involved in neuroblastoma, a pediatric cancer involving nerve tissue (Supplementary \ref{supp:S1}). We have  docked 2,584 FDA-approved small molecules to AlphaFold-predicted human protein structures using both PocketVina and CompassDock \cite{sarigun2024compassdock}. The resulting protein–ligand pairs were cross-referenced with PubChem \cite{kim2016pubchem} bioactivity annotations. Each pair was labeled as “active” or “inactive” based on known experimental outcomes, yielding 1,446 active and 15,211 inactive complexes from a total of 563,251 drug-target pairs (not all pairs have bioactivity annotations in PubChem).

\vspace{1em}

We compared PocketVina with CompassDock, a protocol that combines DiffDock-L-generated ligand conformations with semi-empirical scoring for binding affinity (or AA-Score \cite{pan2022aa}) and a physicochemical validation tool, PoseCheck \cite{harris2023posecheck}. CompassDock evaluates poses using metrics such as AA-Score binding affinity, protein–ligand steric clashes, ligand strain energy, and model confidence scores, layered on top of DiffDock-L’s generated conformations. In contrast, PocketVina directly samples and scores conformations using binding affinity values returned by the QuickVina-based backend.

\vspace{1em}

\begin{figure}[h]
\centering
\includegraphics[width=0.99\textwidth]{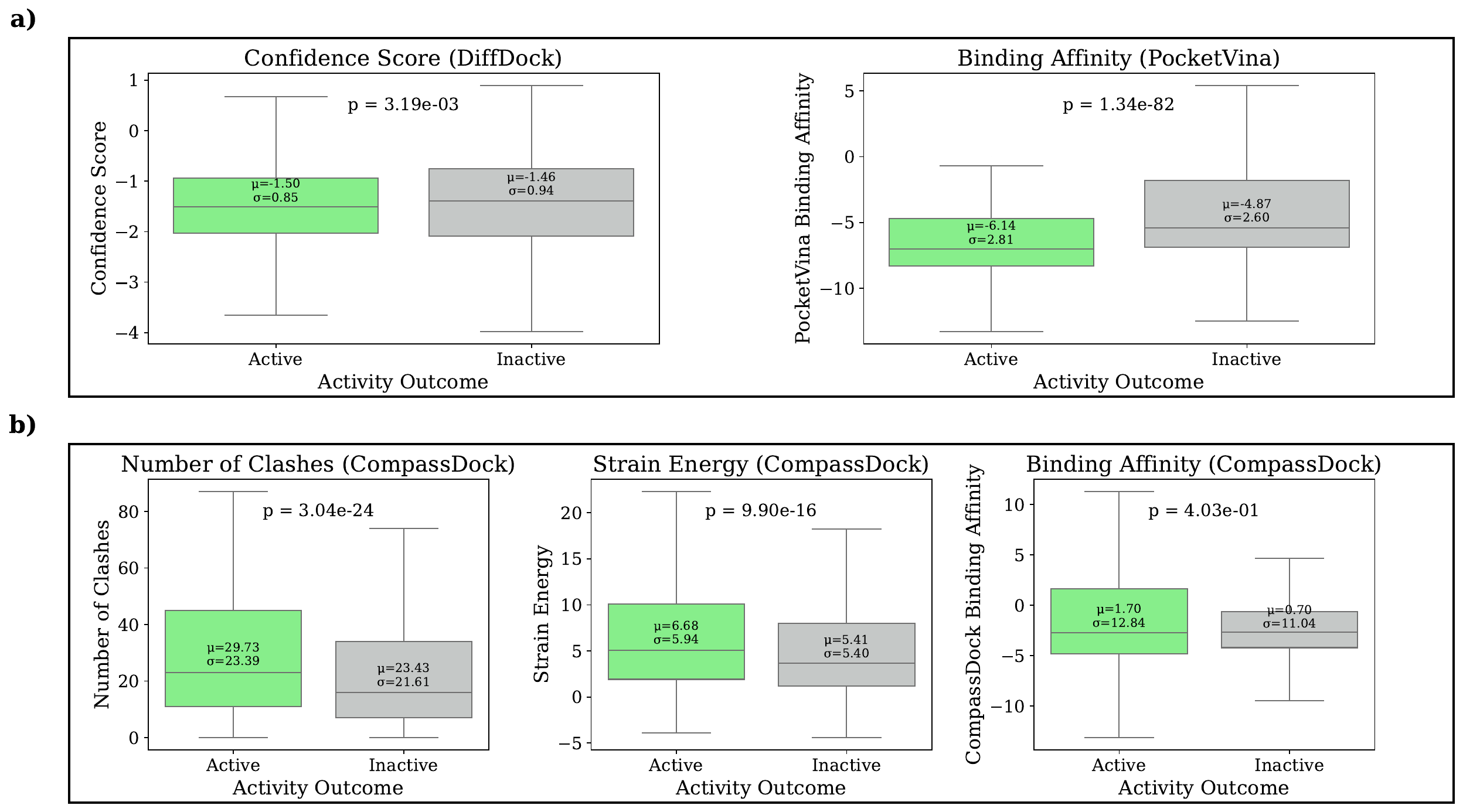}
\caption{%
  \textbar{} \textbf{Comparison of Active and Inactive Drug–Target discriminations with docking methods on the TargetDock-AI (AI = Active/Inactive) dataset.}
  \newline
  Box-and-whisker plots compare five docking-derived metrics for compounds labeled ``Active'' versus ``Inactive'': 
  \textbf{(a)} DiffDock-L confidence score (left) and PocketVina binding affinity (right; computed via QuickVina). 
  \textbf{(b)} Post-processed metrics from CompassDock, evaluated on top-ranked poses from DiffDock-L: number of protein–ligand steric clashes (left; computed via PoseCheck), strain energy (middle; PoseCheck), and binding affinity (right; computed using AA-Score).
  \newline
  Each box represents the interquartile range (IQR), defined as the distance between the 75th (Q3) and 25th (Q1) percentiles of the data, with the horizontal line indicating the median (Q2). 
  The mean (\(\mu\)) and standard deviation (\(\sigma\)) are also annotated on each panel. 
  Statistical significance between ``active'' and ``inactive'' groups was assessed using two-sided Wilcoxon rank-sum tests; p-values are shown above each comparison.
}
\label{figure_5}
\end{figure}

As shown in Figure \ref{figure_5}, only PocketVina (panel a, right) was able to meaningfully distinguish between “active” and “inactive” drug–target pairs using the QuickVina scoring function, with a highly significant separation (\(p = 1.34 \times 10^{-82}\)). In contrast, none of the metrics evaluated within CompassDock (panel a, left, and panel b)—including AA-Score binding affinity (\(p = 0.403\); lower is better), number of steric clashes (\(p = 3.04 \times 10^{-24}\); lower is better), strain energy (\(p = 1.71 \times 10^{-4}\); lower is better), and DiffDock-L confidence score (\(p = 3.19 \times 10^{-3}\); higher is better)—showed both directionally consistent and biologically meaningful separation between the two groups. Notably, some of these metrics (e.g., steric clashes, strain energy and confidence score) produced statistically significant differences, but in the wrong direction—highlighting strong false positive signals. These findings reinforce a key limitation: the effectiveness of a scoring function is closely tied to the quality of the underlying pose generation \cite{ricci2021improving, tran2023practical}.

\vspace{1em}

These results suggest that even without training on activity data, PocketVina’s combination of multi-pocket conditioned sampling and Vina-based scoring provides sufficient discriminatory power to prioritize likely bioactive compounds. This makes it a promising tool for large-scale virtual screening workflows, particularly in early-stage discovery, where predicted binding affinity remains a practical proxy for biological activity \cite{higham2024affinity}.

\vspace{1em}

As shown in Table \ref{table_3}, PocketVina successfully distinguishes between “active” and “inactive” drug–target pairs within approximately three days, using seven GPUs and $\sim$6 GB of VRAM. In comparison, CompassDock required approximately 1.5 months to complete the same benchmark, utilizing $\sim$20 GPUs and $\sim$15 GB of VRAM. A substantial portion of the total runtime in CompassDock arose from downstream evaluation steps—including steric clash detection, ligand strain energy estimation, and AA-Score computation—performed on top-ranked poses generated by DiffDock-L.

\vspace{1em}

Although DiffDock-L’s internal confidence scores offer a ranking mechanism for sampled poses, prior work \cite{buttenschoen2024posebusters} has shown that such scores do not reliably capture physical or chemical plausibility. To address this, CompassDock applies additional physicochemical validation steps—including steric clash checks, strain energy calculations, and AA-Score evaluation—to improve the interpretability and reliability of its predictions. These post-processing steps, however, come with significant computational overhead. In contrast, PocketVina’s integrated approach couples multi-pocket sampling with Vina-based affinity scoring, enabling efficient and accurate large-scale virtual screening.

\begin{table}[h]
  \centering
  \caption{\textbar{} Summary of docking methods’ performance on the TargetDock-AI (AI = Active/Inactive) dataset}
  \label{table_3}
  \begin{tabular*}{\textwidth}{@{\extracolsep\fill} l c c @{}}
    \toprule
    & \textbf{\makecell{CompassDock\\(DL-Based)}} 
    & \textbf{\makecell{PocketVina-API\\(Search-Based)}} \\
    \midrule
    \textbf{Runtime} & $\sim$1.5 months & $\sim$3 days  \\
    \textbf{GPU-VRAM}  & $\sim$15\,GB     & $\sim$6\,GB   \\
    \textbf{\# of GPUs}  & 20   & 7   \\
    \textbf{Discriminative for Actives/Inactives} & \xmark  & \cmark \\
    \bottomrule
  \end{tabular*}
\end{table}

\section*{Discussion}\label{sec3}

\vspace{1em}

\hspace*{1.0em} DL-based and traditional ligand docking methods continue to face challenges in achieving high rates of physically valid docking. In this study, we present PocketVina, a multi-pocket conditional docking system to improve the quality of sampled binding conformations.

\vspace{1em}

We analyzed PocketVina against various benchmarks, where PocketVina consistently outperformed both DL-based and traditional methods in terms of physically valid docking success rate, and continued to outperform other methods on novel proteins. When we analyzed the effect of ligand flexibility on the success rate, PocketVina still maintained the highest physically valid docking success rate, despite a significant decrease in valid success rate for flexible ligands. Furthermore, PocketVina was not only able to accurately distinguish between “active” and “inactive” drug-target pairs, but also demonstrated highly efficient runtime performance. These features strengthen its potential for hit identification on large chemical libraries. Overall, our results confirm that PocketVina can be effectively applied in real-world scenarios with its multi-pocket-conditioned ligand docking capability.

\vspace{1em}

While PocketVina shows strong performance in sampling physically valid binding conformations, the relationship between pocket rank and Vina-based binding affinity, when assessed on best poses, does not consistently yield a clear ranking. This observation aligns with the intended scope of PocketVina, which focuses on improving conformational sampling rather than redefining scoring strategies. Notably, while PocketVina does not aim to optimize scoring, this analysis suggests that limiting docking to the top five pockets is sufficient for effective virtual screening. This approach offers a favorable trade-off between accuracy and runtime, making it well-suited for large-scale applications.

\vspace{1em}

The ability to accurately predict protein-ligand complexes could accelerate the design of novel therapeutic agents by deepening our understanding of protein biology. Given the advances in computational resources and increased access to different chemical spaces, we aim to make PocketVina a key tool in the drug discovery community through our continued development.

\section*{Methods}\label{sec4}

\subsection*{Pocket Prediction with P2Rank
}\label{sec4.1}

\vspace{1em}

\hspace*{1.0em} To identify candidate ligand-binding regions on the protein surface, we used P2Rank \cite{krivak2018p2rank, polak2025prankweb}, a machine learning–based method that predicts ligandable pockets based on geometric and physicochemical properties of the solvent-accessible surface (SAS). In the following, we provide a technical description of the underlying P2Rank, which is used in our workflow without modification.

\vspace{1em}

The method begins by generating a regularly spaced set of surface points $ p \in \mathcal{P}_{\text{SAS}} $ on the SAS, computed using a numerical algorithm \cite{eisenhaber1995double} from the CDK library \cite{steinbeck2003chemistry}. For each surface point $ p $, a feature vector $ \mathbf{f}_p $ is constructed by aggregating information from nearby atoms $ a \in \mathcal{N}_p $ within a 6 Å neighborhood. Specifically, for each atomic property $ \phi $ (e.g., partial charge, hydrophobicity), a distance-weighted sum is calculated:

\vspace{1em}

\begin{equation}
f_p^\phi = \sum_{a \in N_p} \max\bigl(0,\,1 - \tfrac{\mathrm{dist}(a,p)}{6.0}\bigr)\,\phi(a)
\label{prank_eq_1}
\end{equation}

\vspace{1em}

\noindent Additional geometric descriptors such as protrusion—quantified as the number of atoms within 10 Å of the surface point—are also appended to $ \mathbf{f}_p $.

\vspace{1em}

\noindent The resulting feature vector is then passed to a Random Forest model $ \mathcal{M}_{\text{RF}} $, which assigns a ligandability score $ s_p = \mathcal{M}_{\text{RF}}(\mathbf{f}_p) $ to each point. Points with scores exceeding a threshold $ \theta $ are grouped into clusters using single-linkage clustering with a 3 Å cutoff. Each cluster $ \mathcal{K} $ is scored by summing the squared ligandability scores of its constituent points:

\vspace{1em}

\begin{equation}
S_{\mathcal{K}} = \sum_{p \in \mathcal{K}} s_p^2
\label{prank_eq_2}
\end{equation}

\vspace{1em}

\noindent The geometric center of each cluster defines the predicted pocket center, and neighboring residues are identified. All predicted pockets are then ranked by $ S_{\mathcal{K}} $. This overall procedure is summarized in Supplementary Algorithm \ref{supp_alg_1}, which details the computational flow from surface point generation to pocket ranking.

\subsection*{Accelerated Docking with QuickVina 2-GPU 2.1}\label{sec4.2}

\vspace{1em}

\hspace*{1.0em} We used QuickVina 2-GPU 2.1, an OpenCL-based molecular docking framework optimized for high-throughput virtual screening (HTVS) \cite{tang2024vina}. This method builds upon Vina-GPU 2.0 \cite{ding2023vina} by incorporating grid caching, memory-efficient optimization, and OpenCL-level parallelism. The optimization routine, known as Reduced Iteration and Low-Complexity BFGS (RILC-BFGS) improves computational performance by retaining only the most recent iterations, thereby reducing memory usage, communication overhead, and total runtime \cite{zheng2022lomets3}. In the following, we provide a technical description of the underlying QuickVina 2-GPU 2.1, which is used in our workflow without modification.

\vspace{1em}

The docking workflow begins with the generation of a grid cache $ G_0, G_1, \dots, G_T $ over the protein target $ P $, which stores precomputed values on a 3D grid for each ligand atom type. This grid is reused across all ligands, significantly reducing CPU–GPU communication and redundant computations—commonly referred to as the Kernel 1 phase in the pipeline.

\vspace{1em}

In Kernel 2, for each ligand $ L_i $, the algorithm launches $ n $ OpenCL threads, each initialized with a random conformation $ C_i $, represented by its position, orientation, and torsions (POT) as:

\vspace{1em}

\begin{equation}
C_i = \{x^i, y^i, z^i, a^i, b^i, c^i, d^i, \psi_1^i, \psi_2^i, \dots, \psi^i_{N_{\text{rot}}} \}
\label{qvina_eq_1}
\end{equation}

\vspace{1em}

\noindent Here $ x^i, y^i, z^i $ specify the spatial coordinates within the search space; $ a^i, b^i, c^i, d^i $ are quaternion components encoding rigid-body orientation; and $ \psi_k^i $ are torsion angles for each of the $ N_{\text{rot}} $ rotatable bonds. A new conformation $ C_i' $ is generated via uniform random mutation of a POT element:

\vspace{1em}

\begin{equation}
C_i' = \mathcal{R}(C_i)
\label{qvina_eq_2}
\end{equation}

\vspace{1em}

\noindent where $ \mathcal{R}(\cdot) $ is a jitter function that perturbs one POT degree of freedom. Each conformation is evaluated using a scoring function:

\vspace{1em}

\begin{equation}
\text{SF}_{C_i'} = f(C_i') = e_{\text{inter}} + e_{\text{intra}}
\label{qvina_eq_3}
\end{equation}

\vspace{1em}

\noindent where the interaction energy of the paired atoms inside the ligand is defined by the $ e_{\text{intra}} $, while the interaction energy between the ligand and the receptor is denoted by the $ e_{\text{inter}} $. Trilinear interpolation is used to calculate the $ e_{\text{inter}} $ by searching the grid cache.

\vspace{1em}

\noindent Before proceeding to local optimization, a first-order consistency check is applied to determine whether the current conformation warrants refinement. This check evaluates the sign of the scoring function’s partial derivatives across neighboring conformations $ C_j $:

\vspace{1em}

\begin{equation}
\text{sign} \left( \left.\frac{\partial \text{SF}_C}{\partial C_p} \right|_{C = C_i'} \right) \cdot \text{sign} \left( \left.\frac{\partial \text{SF}_C}{\partial C_p} \right|_{C = \bar{C}_j} \right) \leq 0
\label{qvina_eq_4}
\end{equation}

\vspace{1em}

\noindent where $  \left.\frac{\partial \text{SF}_C}{\partial C_p} \right|_{C = m}  $ is the partial derivative of the scoring function $ \text{SF} $ with respect to the $ C_p $ at the point $ \text{m}$, and $ \bar{C}_j $ denotes a neighboring conformation of $ C_i' $ in the POT space, sampled for derivative consistency checks. $ \text{sign}(\cdot) $ is sign function which is an odd mathematical function that extracts the sign of a real number. If this fails, a secondary geometric-energetic condition is evaluated:

\vspace{1em}

\begin{equation}
\text{sign} \left( \left.\frac{\partial \text{SF}_C}{\partial C_p} \right|_{C = C_i'} \right) \cdot \text{sign} \left\{ \left[\text{SF}_{C_i'} - \text{SF}_{\bar{C}_j} \right] \cdot \left[ (C_i')_p - (\bar{C}_j)_p \right] \right\} \leq 0
\label{qvina_eq_5}
\end{equation}

\vspace{1em}

\noindent If the conformation passes the check, local optimization is performed using a RILC-BFGS method, which improves upon classical BFGS by introducing early stopping, robust line search, and memory-efficient updates tailored for GPU execution.

\vspace{1em}

\noindent The optimization proceeds iteratively, where at each iteration $ k $, the conformation vector is denoted as $ \mathbf{x}_k $, and the corresponding gradient of the scoring function as $ \nabla \text{SF}(\mathbf{x}_k) $. The descent direction is initialized as:

\vspace{1em}

\begin{equation}
\mathbf{d}_k = -\nabla \text{SF}(\mathbf{x}_k)
\label{qvina_eq_6}
\end{equation}

\vspace{1em}

\noindent To determine the step size $ \lambda_k $, RILC-BFGS uses a Lewis–Overton inexact line search, which seeks a value satisfying both the Armijo condition:

\vspace{1em}

\begin{equation}
\text{SF}(\mathbf{x}_{k+1}) > \text{SF}(\mathbf{x}_k) + \lambda_k \rho \, \nabla \text{SF}(\mathbf{x}_k)^\top \mathbf{d}_k
\label{qvina_eq_7}
\end{equation}

\vspace{1em}

\noindent and the weak Wolfe condition:

\vspace{1em}

\begin{equation}
\nabla \text{SF}(\mathbf{x}_{k+1})^\top \mathbf{d}_k < \sigma \, \nabla \text{SF}(\mathbf{x}_k)^\top \mathbf{d}_k
\label{qvina_eq_8}
\end{equation}

\vspace{1em}

\noindent Here, $ \rho $ and $ \sigma $ are standard line search parameters (e.g., $\rho = 10^{-4} $, $ \sigma = 0.1 $). Once a valid step size is found, the conformation is updated as:

\vspace{1em}

\begin{equation}
\mathbf{x}_{k+1} = \mathbf{x}_k + \lambda_k \mathbf{d}_k
\label{qvina_eq_9}
\end{equation}

\vspace{1em}

\noindent The displacement and gradient difference vectors are then computed:

\vspace{1em}

\begin{equation}
\mathbf{s}_k = \mathbf{x}_{k+1} - \mathbf{x}_k, \quad \mathbf{y}_k = \nabla \text{SF}(\mathbf{x}_{k+1}) - \nabla \text{SF}(\mathbf{x}_k)
\label{qvina_eq_10}
\end{equation}

\vspace{1em}

\noindent To ensure numerical stability, the update is applied only when the following curvature condition is satisfied:

\vspace{1em}

\begin{equation}
\frac{\mathbf{y}_k^\top \mathbf{s}_k}{\|\mathbf{s}_k\|^2} > \varepsilon \cdot \|\nabla \text{SF}(\mathbf{x}_k)\|
\label{qvina_eq_11}
\end{equation}

\vspace{1em}

\noindent If the condition holds, the descent direction is refined using a two-loop recursion over the $ m $ most recent pairs ($ \mathbf{s}_i, \mathbf{y}_i $), eliminating the need to explicitly form or store the Hessian. This reduces the computational complexity from $ \mathcal{O}(n^2) $ in classical BFGS to $ \mathcal{O}(mn) $, where $ m \ll n $.

\vspace{1em}

\noindent The RILC-BFGS procedure terminates either when the gradient norm falls below a threshold,

\vspace{1em}

\begin{equation}
\|\nabla \text{SF}(\mathbf{x}_{k+1})\| \le \varepsilon, \quad 0 < \varepsilon \le 1
\label{qvina_eq_12}
\end{equation}

\vspace{1em}

\noindent or when the direction $ \mathbf{d}_k $ ceases to be a descent direction, i.e., $ \nabla \text{SF}(\mathbf{x}_k)^\top \mathbf{d}_k \ge 0 $, at which point the algorithm exits early. After each iteration, the energy before and after mutation is compared. If the post-optimization energy $ e_o $ is worse than the pre-optimization energy $ e_i $, a Metropolis acceptance criterion is applied:

\vspace{1em}

\begin{equation}
P \;=\;
\begin{cases}
  1,                                        & e_i > e_o,            \\  
  \dfrac{\exp(e_i - e_o)}{1.2},             & e_i \le e_o\;.       
\end{cases}
\label{qvina_eq_13}
\end{equation}

\vspace{1em}

\noindent If $ e_o $ is lower than the energy of the current optimal conformation, the structure undergoes further refinement via RILC-BFGS. The algorithm then proceeds to the next iteration, using the newly optimized conformation as the updated starting point.

\vspace{1em}

\noindent The iterative search proceeds until convergence or the maximum number of iterations is reached. Each thread outputs its best pose. The collected conformations are clustered and sorted, followed by rescoring and r.m.s.d.-based filtering. The $ \text{top}-k_i $ poses per ligand are then returned as final output. This workflow is summarized in Supplementary Algorithm \ref{supp_alg_2} and visualized in Supplementary Figure \ref{Supp_figure_3}.

\subsection*{PocketVina}\label{sec4.3}

\hspace*{1.0em} PocketVina follows a two-step procedure. It first applies P2Rank to the input protein structure $ P $ to generate a set of predicted binding pockets. It then extracts the center coordinates of these pockets as $ \{c_1, c_2, \dots, c_k\} $ (Supplementary Algorithm \ref{supp_alg_1}).

\vspace{1em}

For each pocket center $ c_i $, PocketVina constructs a user-defined docking search box and runs QuickVina 2-GPU 2.1 to dock the ligand $ L $ into the corresponding region. During docking, QuickVina generates multiple conformations $ \{C_{i,1}, \dots, C_{i,n}\} $ and computes binding affinity scores $ E(C_{i,j}) $ using its internal scoring function (Supplementary Algorithm \ref{supp_alg_2}).

\vspace{1em}

PocketVina aggregates the docking results from all pocket centers and outputs the final set of conformations along with their predicted affinities. PocketVina is summarized in Supplementary Algorithm \ref{supp_alg_3} and visualized in Figure \ref{figure_1}.

\section*{Data Availability}\label{sec5}

The PoseBusters, Astex Diverse, DockGen, and PDBbind2020 datasets used in this study are publicly available and described in Supplementary \ref{supp:S1}. The TargetDock-AI dataset was curated as part of this work, as detailed in Supplementary \ref{supp:S1}. All benchmark results, including docking outputs and evaluation metrics, are available \emph{via} Zenodo at \url{https://zenodo.org/records/15733460}. The TargetDock-AI dataset is also accessible at the same repository.

\section*{Code Availability}\label{sec6}

The source code used to perform the PocketVina docking experiments described in Supplementary \ref{supp:S2} is available \emph{via} GitHub under the MIT License at \url{https://github.com/BIMSBbioinfo/PocketVina}.

\section*{Acknowledgements}\label{sec7}

A.S. and A.A. are supported by the Helmholtz IVF fund. The experiments were carried out using the MDC-HPC (Max-Cluster), as well as AkalinLab-Hulk and AkalinLab-Beast servers. We extend our sincere thanks to Dr. Martin R. Siegert (MDC), Dan Munteanu (MDC), Madalin Ionel Patrascu (MDC), and Ricardo Wurmus (MDC) for their support throughout this work. 

\section*{Competing Interests}\label{sec8}

The authors declare no competing interests.

\section*{Authors Contributions}\label{sec9}

A.A. conceived and planned the project together with A.S. A.S. curated the datasets, implemented the software, designed and performed the experiments, conducted data analysis, created the visualizations, and prepared the figures and manuscript. B.U. and V.F. contributed to the interpretation of results and provided critical revisions to the manuscript. A.A. supervised the work, guided data interpretation and experimental design, edited the manuscript, and provided funding and resources. All authors approved the final version of the manuscript.

\vspace{1em}

\newpage

\bibliography{sn-bibliography}

\newpage

\makeatletter
  \setcounter{figure}{0}

  \renewcommand{\figurename}{Extended Data Fig.}

  \renewcommand{\thefigure}{E\arabic{figure}}

  \renewcommand{\theHfigure}{E\arabic{figure}}
\makeatother

\begin{figure}[!htbp]
\centering
\includegraphics[width=0.96\textwidth]{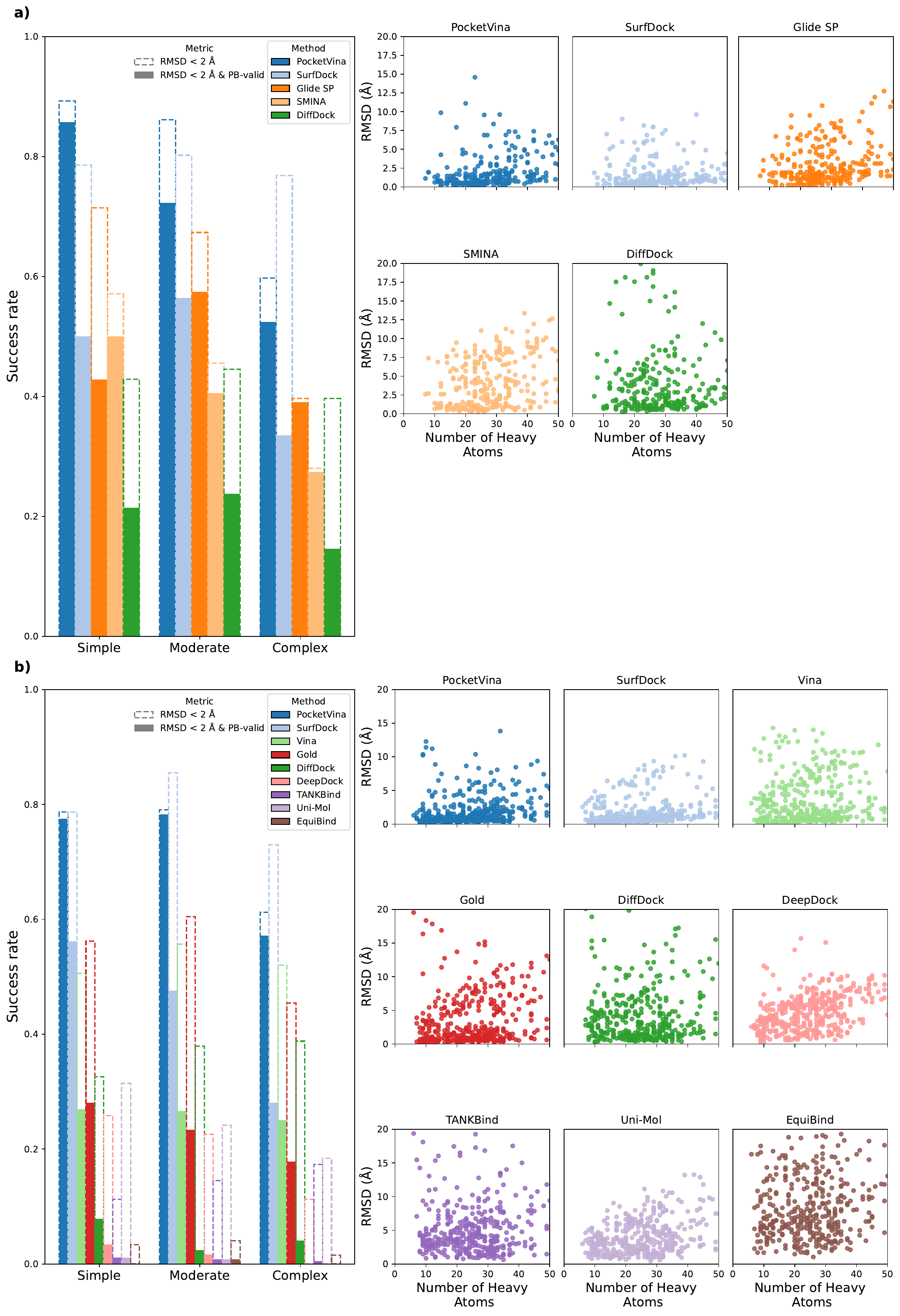}
\caption{\textbar{} \textbf{Physically valid docking performance across ligand complexity levels.} (simple: $\leq$ 15 heavy atoms; moderate: 15–25; complex: $>$ 25) 
\newline
\textbf{(a)} Effect of heavy atoms on physically valid docking success rates and r.m.s.d. distribution in the PDBbind time-split set.
\textbf{(b)} Effect of heavy atoms on physically valid docking success rates and r.m.s.d. distribution in the PoseBusters set.
}
\label{Ext_figure_1}
\end{figure}

\vspace{1em}

\begin{figure}[h]
\centering
\includegraphics[width=0.96\textwidth]{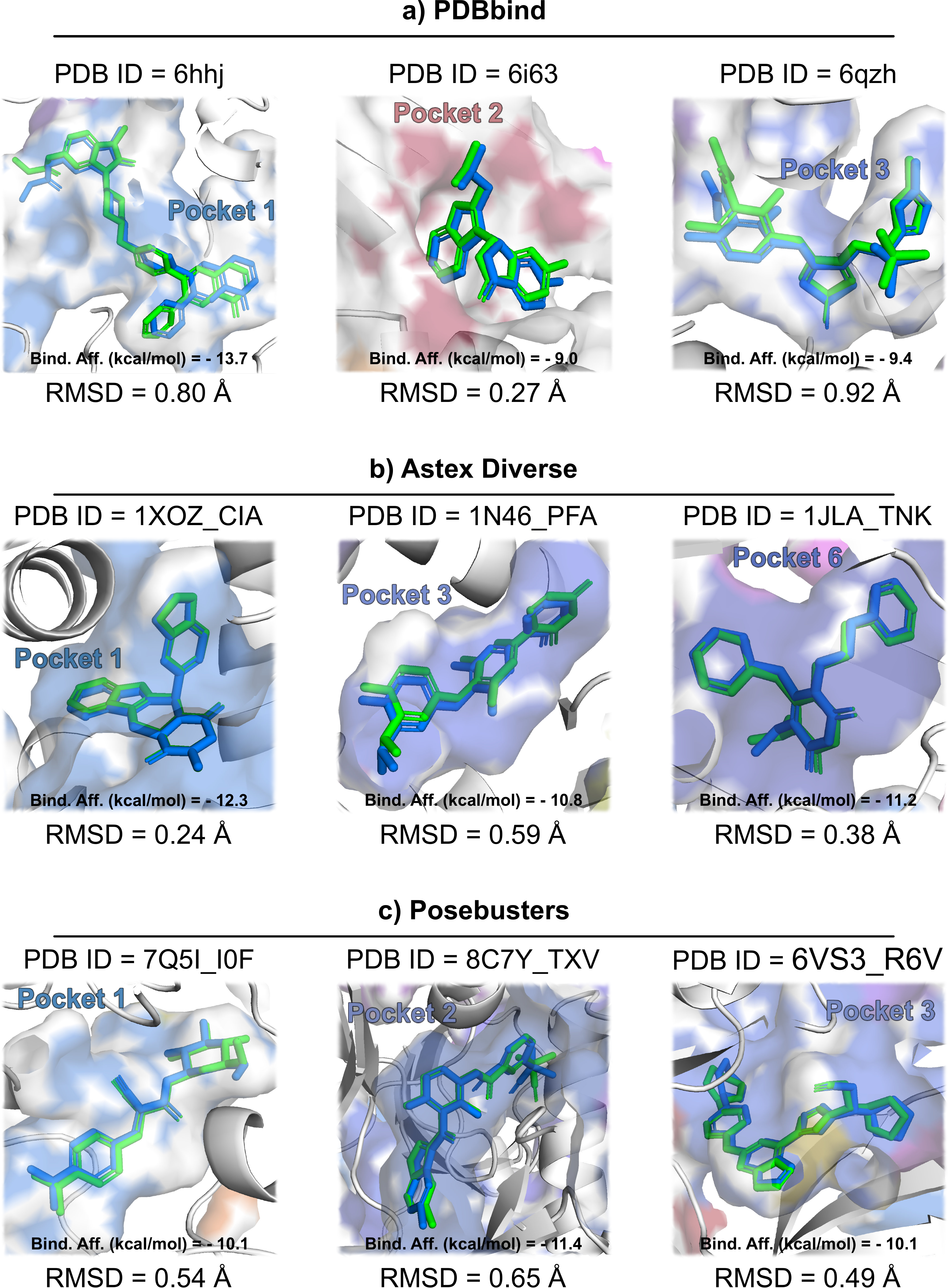}
\caption{\textbar{} \textbf{Visualization of ligand poses in pocket region sampled by PocketVina.}
\newline
Crystal ligand poses are shown in \textcolor{mygreen}{green}, and PocketVina-predicted poses (best r.m.s.d.) are shown in \textcolor{blue}{marine blue}. Pocket regions are colored according to their predicted ligandability scores from P2Rank. Visualizations were generated using the PyMOL \cite{delano2002pymol} script provided by P2Rank. For each ligand, the corresponding pocket rank and predicted binding affinity (from QuickVina 2-GPU 2.1) are shown. \textbf{(a)} Representative poses from the PDBbind2020 dataset. \textbf{(b)} Representative poses from the Astex Diverse dataset. \textbf{(c)} Representative poses from the PoseBusters dataset.
}
\label{Ext_figure_2}
\end{figure}

\clearpage

\renewcommand{\thesection}{S\arabic{section}}

\setcounter{section}{0}

\section*{\centering\huge Supplementary Materials}
\addcontentsline{toc}{section}{Supplementary Materials}

\vspace{1em}
\vspace{1em}

\section{Datasets and Literature Results}\label{supp:S1}

\subsection*{PDBbind}
\vspace{1em}

We employed the PDBbind2020 time-split dataset as a benchmark, as it is commonly used in deep learning–based docking studies. This benchmark comprises a diverse collection of compounds, including both small molecules and peptides, making it well-suited for evaluating docking performance across different ligand types. We obtained results for EquiBind, TANKBind, E3Bind, and Uni-Dock from Ke \emph{et al.} \cite{yu2023deep}; KarmaDock from its original paper \cite{zhang2023efficient}. DiffDock’s blind-docking results were retrieved from the open-source repository maintained by Ke \emph{et al.} \cite{yu2023deep} (\url{https://github.com/pkuyyj/Blind_docking}), while its pocket-conditioned results follow Huang \emph{et al.} \cite{huang2024re}. Glide SP, GNINA, SMINA, Vina, DiffDock-L and SurfDock were obtained from SurfDock \cite{cao2025surfdock}.

\subsection*{PoseBusters and Astex Diverse}
\vspace{1em}

The PoseBusters benchmark, consisting of 428 complexes involving drug-like molecules released after 2021, serves as an independent test set. Since DL models trained on PDBbind2020 have not been exposed to these structures, it offers a reliable measure for assessing model generalization to unseen data.

\vspace{1em}

\noindent We additionally evaluate on the Astex Diverse Set, a widely used a benchmark introduced in 2007, whose entries largely overlap with those in the PDBbind2020 training set. For consistency and comparability, we obtained performance results for Vina, Gold, DiffDock, DeepDock, TANKBind, Uni-Mol and EquiBind from the PoseBusters study \cite{buttenschoen2024posebusters}. while KarmaDock, DiffDock-L, SurfDock(unminimized), Smina, GLIDE SP and GNINA from SurfDock paper \cite{cao2025surfdock}.

\subsection*{DockGen}
\vspace{1em}

To address the issue of structural redundancy in existing docking benchmarks, Corso \emph{et al.} \cite{corso2024deep} introduced a new dataset, DockGen, highlighting that binding pockets can remain highly similar despite low sequence identity between training and test proteins. We used the DockGen-full testset, consisting of 189 protein–ligand complexes, in our evaluation. We obtained the results for GNINA, DiffDock, DiffDock-L and SurfDock from SurfDock \cite{cao2025surfdock} paper.

\subsection*{TargetDock-AI}
\vspace{1em}

TargetDock-AI is a benchmark dataset that we curated to evaluate whether docking-derived metrics can reliably distinguish between active and inactive drug–target interactions in the absence of task-specific training. Protein–ligand pairs were annotated using bioactivity data from PubChem. Each pair was labeled as “active” if the compound was experimentally confirmed to bind the target, and “inactive” if the compound was tested but showed no activity. The final dataset contains 563,251 protein–ligand pairs in total, among which 1,446 are labeled as active and 15,211 as inactive. The full list of unique gene names represented in the TargetDock-AI dataset is provided below:

\vspace{1em}

\noindent GRM8, CDC7, PLK4, E2F3, KMT2D, CHEK1, TP73, XRCC2, DCX, TLX2, SPAG9, KDM1A, BUB1B, SHOX2, SAP30, PIAS2, CCNB2, SNAI1, HAND1, NSD2, KRAS, INS, NTRK1, TP53, CHGB, FYN, CDK1, CRH, TH, NEFL, RET, ABCB1, SYP, ITGA2B, DBH, GNAO1, PARP1, MYB, MYBL2, BCL2, MAPT, TOP2A, ODC1, SRC, NCAM1, MTHFD2, CCNB1, POU2F1, BRAF, MYOD1, CHN1, TCF3, CREB1, DDX5, GABRG2, RCC1, CSNK2A2, SYT1, GAL, GLDC, PAX7, GATA2, GATA3, SLC6A2, CCNE1, MCM3, YY1, LMO1, DNMT1, APEX1, CHAT, WEE1, GNRHR, TLX1, AKT2, L1CAM, ARRB2, CHRNA3, MCM4, MCM5, MCM7, BMI1, HOXD13, GRK3, SOX11, NUP62, BRCA1, FEN1, CCNF, PRPH, SLC19A1, CASP2, SLC1A2, DCC, ETV4, GATA4, MAPK8, CRKL, NSF, GCLC, NES, PRLHR, CTCF, MCM2, GSK3B, MRE11, MNX1, SMARCA4, HCFC1, USP11, MSH6, MAPK10, ALDH18A1, UCP3, DLX5, CTBP2, WNT3, HSD17B7, INHBE, SST, WDR5, PPP1CA, UBE2I, YBX1, CSNK2A1, CDK6, MDM2, E2F1, INSM1, SATB1, POU4F1, SP3, PAX5, PTK2, PRKCZ, RAD51, SOX4, SREBF2, POU4F2, GRIK3, MAP2K5, MAPK7, TBX2, TRIM28, SKP2, CDK5R2, MTA1, CTBP1, LSAMP, SMAD4, PIN1, PTCH1, E2F2, MCM6, SUZ12, POLD3, CDK5R1, E2F5, EPHA7, MED1, EFNB3, TBCE, EZH2, NTRK3, DDB1, CCNG2, FSCN1, SGO1, TET2, LRRN1, BCOR, LIN28B, TUBA1A, SV2A, MCM10, LDB1, SULF2, WWC1, ATOH7, SFRP1, GPR161, CAMKV, DOT1L, NUP210, DEPDC1B, LMBR1, SORCS1, DDX1, AKAP1, HDAC2, FBXW7, RMI2, SIRT1, GALNT14, AURKB, SMC6, MMS19, EYA1, CDC6, SMO, VRK1, WDR77, FSD1, CALN1, PECR, NSD3, NUF2, FOXP3, BCL11B, SYT4, SMARCAD1, SMYD3, SLC25A19, EML4, SIX2, FZD3, INCENP, DDX21, DLL3, CHD7, NRXN2, ONECUT1, PCSK1N, MCM8, NRXN1, NDRG4, ALK, TBX20, PLAGL2, CTNND2, GAB2, SCML2, ZNF281, HCN4, PRKAB1, TRRAP, KDM3A

\newpage

\section{Docking Setups}\label{supp:S2}

\vspace{1em}

\subsection*{PocketVina}

\textbf{Packages and Software:}\\

\vspace{0.05em}

\noindent QuickVina 2-GPU 2.1, P2Rank 2.5, Boost 1.77, RDKit 2024.9.6, Open Babel 3.1.1.21, Numpy 2.2.4, Pandas 2.2.3.

\vspace{1em}

\noindent \textbf{Parameters:}\\

\vspace{0.05em}

\noindent We used the pocket prediction step using three P2Rank configurations: the “\textnormal{default}” model, “\textnormal{alphafold}” model, and “\textnormal{alphafold\_conservation}” model. Unless otherwise specified, the results reported in the main text correspond to the “\textnormal{alphafold}” P2Rank model, as it yielded robust predictions across diverse targets.

\vspace{1em}

\noindent For docking with QuickVina 2-GPU 2.1, we used mode 20, which performs sampling of 20 ligand conformations per pocket center. To account for different possible binding region sizes, we tested a range of search box sizes from 15 Å to 30 Å, increasing in 1 Å increments. The box was centered at each pocket center predicted by P2Rank. These settings were chosen to balance coverage of plausible binding regions with computational efficiency, and were kept consistent across all benchmark datasets.

\vspace{1em}

\noindent For the TargetDock-AI dataset, we used the “\textnormal{default}” P2Rank model and a fixed box size of 20 Å to standardize the docking procedure across a large number of protein–ligand pairs.

\vspace{1em}

\noindent \textbf{GPU-Device:}\\

\noindent NVIDIA Tesla P40 (for TargetDock-AI)\\
\noindent NVIDIA Tesla T4 (for PoseBusters, Astex, PDBbind2020, and DockGen)

\vspace{1em}

\subsection*{CompassDock}

\textbf{Packages and Software:}\\

\vspace{0.05em}

\noindent CompassDock 0.1.4, DiffDock-L, PoseCheck, AA-Score.

\vspace{1em}

\noindent \textbf{Parameters:}\\

\vspace{0.05em}

\noindent We used open-source codes and weights at \url{https://https://github.com/BIMSBbioinfo/CompassDock}  with default settings.

\vspace{1em}

\noindent \textbf{GPU-Device:}\\

\noindent NVIDIA Tesla P40 and NVIDIA Tesla T4

\vspace{1em}

\section{PoseBusters Evaluation Criteria}\label{supp:S3}

\vspace{1em}

The PoseBusters framework performs a comprehensive set of physical and chemical validity checks on docked ligands. Specifically, it evaluates:


\begin{itemize}
\item Whether the docked ligand can be successfully loaded.
\item Whether the r.m.s.d. is within 2\,\AA{} of the reference ligand.
\item Whether the ligand passes RDKit’s sanity check.
\item Whether the molecular formula is retained compared to the reference ligand.
\item Whether the bond topology is retained compared to the reference ligand.
\item Whether the \textit{sp}\textsuperscript{3} stereochemistry is retained compared to the reference ligand.
\item Whether the double bond stereochemistry is retained compared to the reference ligand.
\item Whether bond lengths fall within acceptable bounds from RDKit’s Distance Geometry parameters.
\item Whether bond angles fall within acceptable bounds from RDKit’s Distance Geometry parameters.
\item Whether the atoms in aromatic rings remain planar.
\item Whether the atoms in double bonds remain planar.
\item Whether internal atomic clashes exist within the ligand.
\item Whether the energy ratio is within thresholds defined by RDKit Universal Force Field (UFF).
\item Whether clashes occur between the ligand and the protein.
\item Whether clashes occur between the ligand and organic cofactors.
\item Whether clashes occur between the ligand and inorganic cofactors.
\end{itemize}


\noindent Together, these criteria help ensure that each docked structure is chemically consistent, physically plausible, and free of major structural anomalies.

\vspace{1em}

\section{Additional Results on PDBBind, PoseBusters and Astex Benchmark Sets}\label{supp:S4}

\vspace{1em}

\hspace*{1.0em} In Supplementary Fig. \ref{Supp_figure_1}, we showed the cumulative distributions of ligand r.m.s.d. values for all evaluated methods across the PDBbind2020 (Timesplit and Unseen), PoseBusters, and Astex benchmark sets. While SurfDock achieves the strongest overall performance in terms of r.m.s.d., PocketVina shows consistently competitive results across all datasets. Among the search-based (i.e., classical) docking methods, PocketVina delivers some of the most robust performance, particularly on structurally diverse and previously unseen targets, highlighting its strong generalization capability without relying on dataset-specific adaptation.



\makeatletter
  \setcounter{figure}{0}

  \renewcommand{\figurename}{Supplementary Fig.}

  \renewcommand{\thefigure}{S\arabic{figure}}

  \renewcommand{\theHfigure}{S\arabic{figure}}
\makeatother

\begin{figure}[h]
\centering
\includegraphics[width=0.99\textwidth]{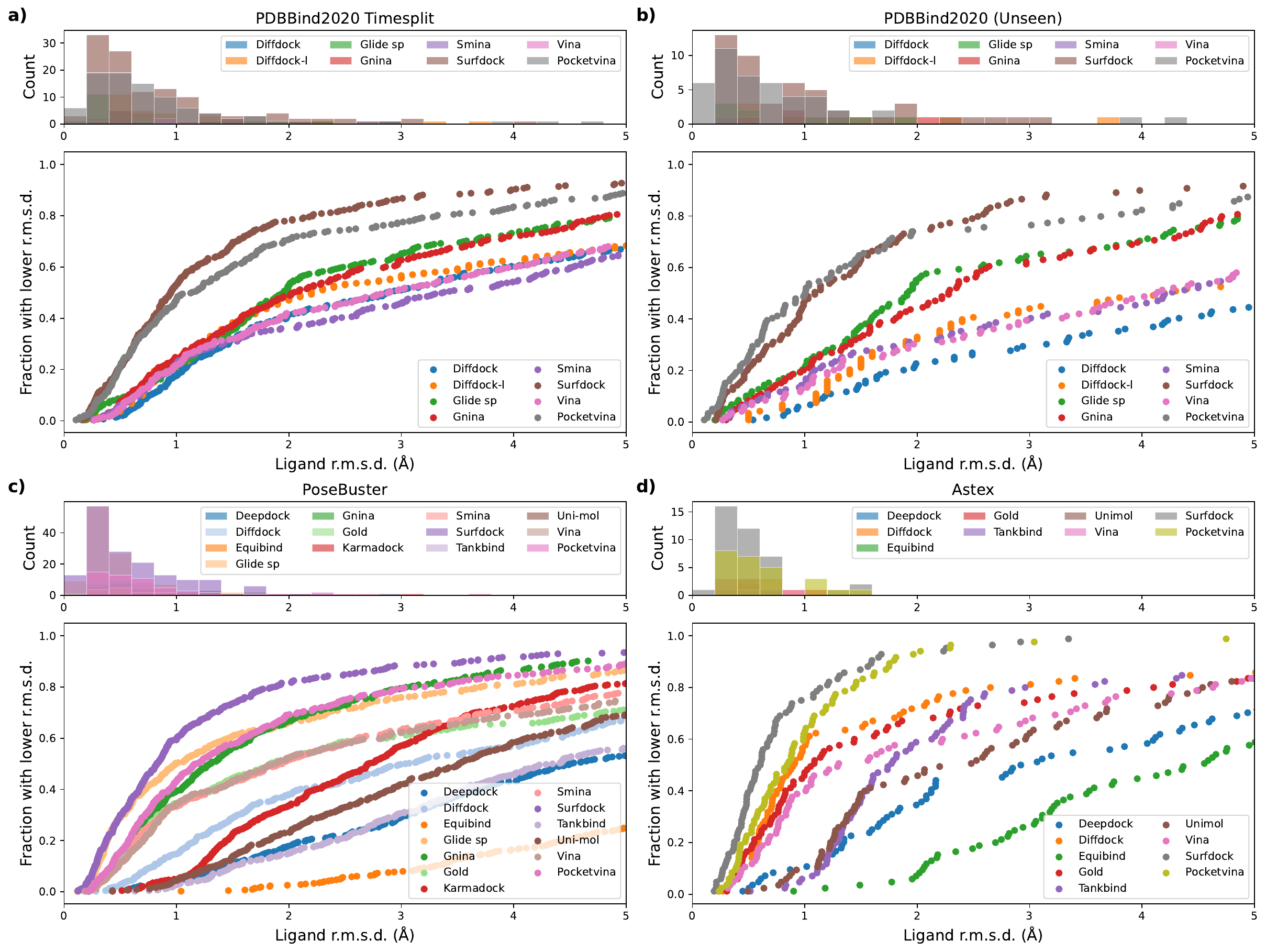}
\caption{\textbar{} \textbf{Cumulative distributions of ligand r.m.s.d. for various docking methods evaluated on the PDBbind (Timesplit and Unseen), PoseBusters, and Astex benchmark sets.} 
\newline
\textbf{(a)} PDBbind2020 Timesplit (363 complexes). \textbf{(b)} PDBbind2020 Unseen (144 complexes). \textbf{(c)} PoseBusters (428 complexes). \textbf{(d)} Astex Diverse (85 complexes).}
\label{Supp_figure_1}
\end{figure}

\vspace{1em}

\section{Ablation Studies of PocketVina on the PDBBind, DockGen, Astex and PoseBusters Benchmark Sets}\label{supp:S5}

\vspace{1em}

In Supplementary Fig. \ref{Supp_figure_2}, we assess the impact of different P2Rank models and box sizes on docking performance across the PDBbind2020, DockGen, Astex, and PoseBusters benchmark sets. We observe that the “alphafold” configuration, particularly when combined with larger box sizes (25 Å and 30 Å), consistently yields lower ligand r.m.s.d. values across all datasets. This suggests that broader spatial coverage around predicted pocket centers, when paired with the “alphafold” P2Rank model, can lead to improved pose accuracy.

\begin{figure}[h]
\centering
\includegraphics[width=0.99\textwidth]{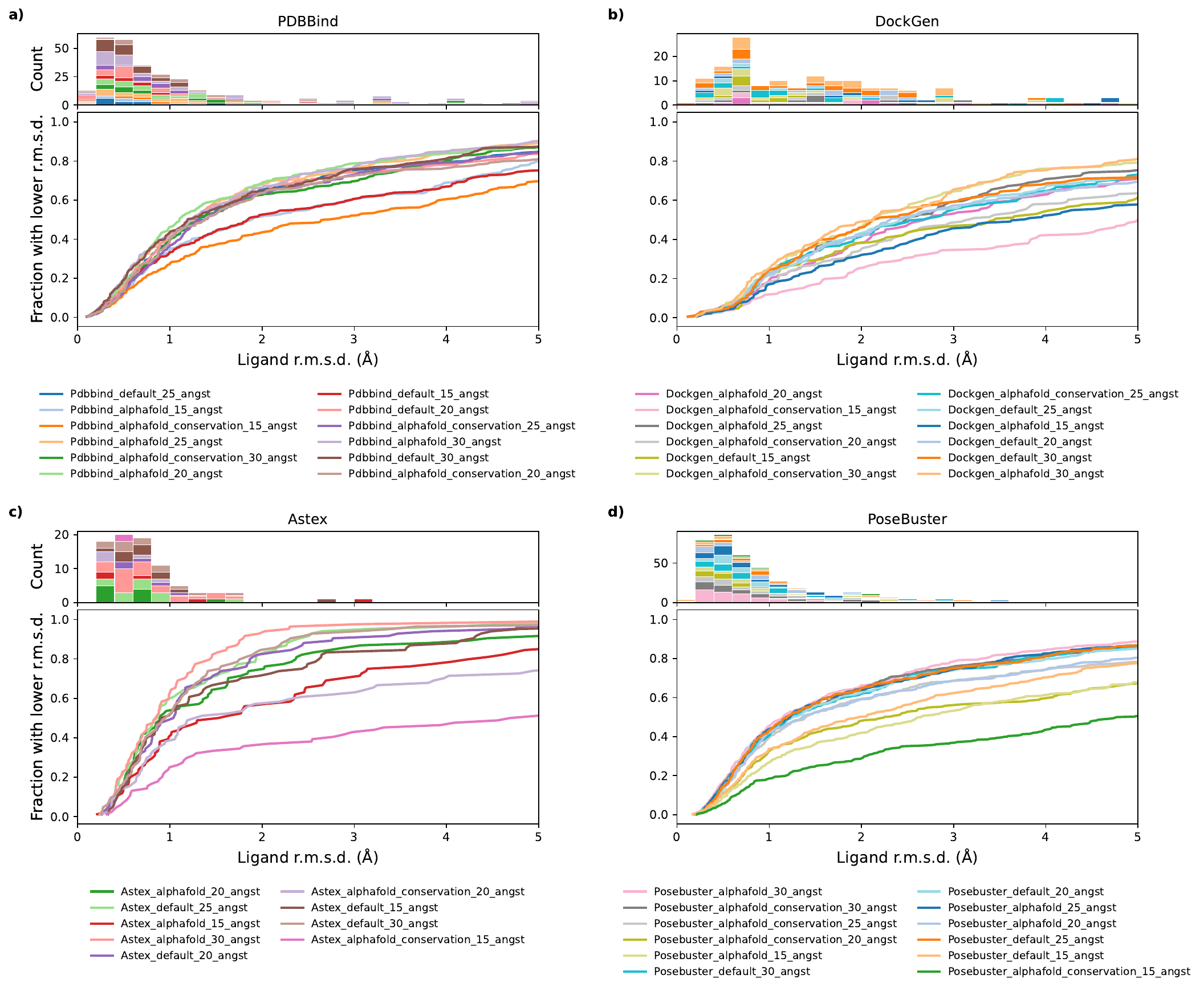}
\caption{\textbar{} \textbf{Ablation study of PocketVina showing cumulative distributions of ligand r.m.s.d. for different P2Rank models and docking box sizes, evaluated on the PDBbind2020, DockGen, Astex, and PoseBusters benchmark sets.} 
\newline
\textbf{(a)} PDBbind2020 Timesplit (363 complexes). \textbf{(b)} DockGen (189 complexes). \textbf{(c)} Astex Diverse (85 complexes). \textbf{(d)} PoseBusters (428 complexes).}
\label{Supp_figure_2}
\end{figure}

\vspace{1em}

\section{Illustration and Pseudocodes for PocketVina Modules}\label{supp:S6}

\vspace{1em}

\begin{figure}[h]
\centering
\includegraphics[width=0.99\textwidth]{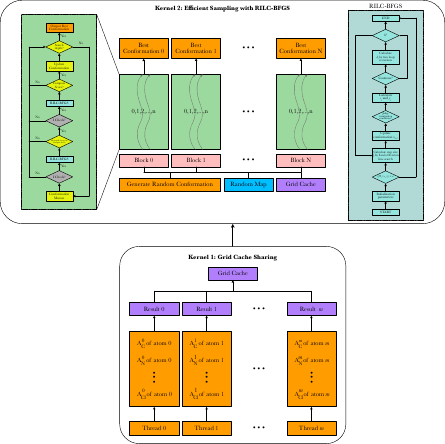}
\caption{\textbar{} \textbf{Overview of QuickVina 2-GPU 2.1} 
\newline
Illustration QuickVina 2-GPU 2.1. The workflow includes ligand and receptor preprocessing via Grid Cache Sharing in Kernel 1, followed by conformation sampling, mutation, consistency checks, RILC-BFGS refinement, Metropolis-based acceptance, and pose selection in Kernel 2.}
\label{Supp_figure_3}
\end{figure}

We provide a visual and algorithmic overview of the main components within the PocketVina workflow. Supplementary Fig. \ref{Supp_figure_3} presents a schematic illustration of the internal workflow of QuickVina 2-GPU 2.1, including conformation initialization, mutation, refinement via RILC-BFGS, and Metropolis-based acceptance. This is followed by three pseudocode listings: Supplementary Algorithm \ref{supp_alg_1} outlines the pocket prediction process using P2Rank; Supplementary Algorithm \ref{supp_alg_2} details the ligand docking procedure implemented in QuickVina 2-GPU 2.1; and Supplementary Algorithm \ref{supp_alg_3} integrates these stages into a unified end-to-end pipeline referred to as PocketVina.

\makeatletter
\renewcommand{\fnum@algorithm}{Supplementary Algorithm~\thealgorithm}
\makeatother

\setcounter{algorithm}{0}

\begin{algorithm}
  \caption{\textbar{} P2Rank Binding Site Prediction}
  \label{supp_alg_1}
  \begin{algorithmic}[1]
    \Procedure{P2Rank\_PredictBindingSites}{protein\_structure}
      \State \textbf{Step 1: Generate surface points}
      \State SAS\_points $\gets$ \Call{GenerateSurfacePoints}{protein\_structure}
      \State
      \State \textbf{Step 2: Compute features for each surface point}
      \ForAll{point \textbf{in} SAS\_points}
        \State neighbors $\gets$ \Call{GetNeighboringAtoms}{point, protein\_structure, 6.0}
        \State feature\_vector $\gets []$
        \ForAll{property \textbf{in} atom\_properties}
          \State value $\gets 0$
          \ForAll{atom \textbf{in} neighbors}
            \State weight $\gets \max\bigl(0,1 - \tfrac{\text{distance}(atom,point)}{6.0}\bigr)$
            \State value $\gets value + weight \times atom.\!get(property)$
          \EndFor
          \State \Call{Append}{feature\_vector, value}
        \EndFor
        \State protrusion $\gets$ \Call{CountAtomsWithin}{point, protein\_structure, 10.0}
        \State \Call{Append}{feature\_vector, protrusion}
        \State point.feature\_vector $\gets$ feature\_vector
      \EndFor
      \State
      \State \textbf{Step 3: Predict ligandability scores}
      \ForAll{point \textbf{in} SAS\_points}
        \State point.score $\gets$ RandomForestModel.\!predict(point.feature\_vector)
      \EndFor
      \State
      \State \textbf{Step 4: Cluster high‐scoring points}
      \State high\_points $\gets \{\,p\in SAS\_points \mid p.score \ge \text{ScoreThreshold}\,\}$
      \State clusters $\gets$ \Call{ClusterPoints}{high\_points, single\_linkage, 3.0}
      \State
      \State \textbf{Step 5: Compute and rank pockets}
      \State pocket\_list $\gets []$
      \ForAll{cluster \textbf{in} clusters}
        \State cluster\_score $\gets \sum_{p\in cluster}(p.score^2)$
        \State center $\gets$ \Call{AverageCoordinates}{cluster.points}
        \State residues $\gets$ \Call{FindResiduesNearPoints}{cluster.points, protein\_structure}
        \State \Call{Append}{pocket\_list, \{score: cluster\_score, center: center, residues: residues\}}
      \EndFor
      \State \Call{SortDescending}{pocket\_list, key=score}
      \State \Return pocket\_list
    \EndProcedure
  \end{algorithmic}
\end{algorithm}

\begin{algorithm}
  \caption{\textbar{} QuickVina 2-GPU 2.1}
  \label{supp_alg_2}
  \begin{algorithmic}[1]
    \Require ligands $\{L_0,\dots,L_M\}$, protein $P$
    \Ensure top $k$ ligand conformations $\{C^*_{i,j}\}$
    \State $G_{0}\dots G_{T} \gets \mathrm{Grid\_cache}(P)$
    \State initialize OpenCL context, queues, kernels; allocate device memory
    \For{$i = 0,\dots,M$} 
      \For{thread $t = 0,\dots,n$}
        \State $C[t]\gets \mathrm{random\_conformation}(L_{i})$
        \State $\mathrm{best}[t]\gets C[t]$
      \EndFor
      \ForAll{thread $t=0,\dots,n$ \textbf{in parallel}}
        \State $C_{\mathrm{cur}}\gets C[t],\ C_{\mathrm{best}}\gets \mathrm{best}[t]$
        \For{$\mathrm{iter}=0,\dots,s'$}
          \State $C_{\mathrm{mut}}\gets \mathrm{Mutate}(C_{\mathrm{cur}})$
          \If{FirstOrderConsistencyCheck($C_{\mathrm{mut}}$)}
            \State $C_{\mathrm{opt}}\gets \mathrm{RILC\_BFGS}(C_{\mathrm{mut}})$
            \If{$\mathrm{Scoring}(C_{\mathrm{opt}})<\mathrm{Scoring}(C_{\mathrm{best}})$}
              \State $C_{\mathrm{best}}\gets C_{\mathrm{opt}}$
            \EndIf
          \EndIf
          \If{Metropolis($C_{\mathrm{cur}},C_{\mathrm{best}}$)}
            \State $C_{\mathrm{cur}}\gets C_{\mathrm{best}}$
          \EndIf
        \EndFor
        \State $\mathrm{best}[t]\gets C_{\mathrm{best}}$
      \EndFor
      \State collect $\{\mathrm{best}[0],\dots,\mathrm{best}[n]\}$; cluster \& sort (GCS)
      \ForAll{selected $C$ in top $k_i$}
        \State refine\_ligand\_structure($C$)
        \State refine\_receptor\_pocket($P,C$)
        \State rescoring($C$)
      \EndFor
      \State apply RMSD\_filtering \& final sort
      \State output top $k_i$ conformations for $L_i$
    \EndFor
    \State \Return $\displaystyle\bigcup_{i=0}^M \{\text{top }k_i\text{ of }L_i\}$
  \end{algorithmic}
\end{algorithm}

\begin{algorithm}
\caption{\textbar{} PocketVina}
\label{supp_alg_3}
\begin{algorithmic}[1]
\Procedure{PocketVina}{receptor\_pdb, ligand\_pdbqt, params}
    \State \textbf{Step 1: Pocket prediction using P2Rank (Supp. Alg. \autoref{supp_alg_1})}
    \State pockets $\gets$ \Call{P2Rank\_PredictBindingSites}{receptor\_pdb}
    \State pocket\_centers $\gets$ \Call{ExtractPocketCenters}{pockets}

    \State \textbf{Step 2: Docking into each pocket using QuickVina 2-GPU 2.1 (Supp. Alg. \autoref{supp_alg_2})}
    \State all\_results $\gets$ []
    \ForAll{center \textbf{in} pocket\_centers}
        \State search\_box $\gets$ \Call{CreateBox}{center, params.box\_size}
        \State poses, Binding\_Affinities $\gets$ \Call{QuickVina 2-GPU 2.1}{receptor\_pdb, ligand\_pdbqt, search\_box, params}
        \State \Call{Append}{all\_results, \{PocketRanks, Binding\_Affinities\}}
    \EndFor


\EndProcedure
\end{algorithmic}
\end{algorithm}

\end{document}